\newcommand{\bq}{\begin{equation}}
\newcommand{\eq}{\end{equation}}
\newcommand{\bqa}{\begin{eqnarray}}
\newcommand{\eqa}{\end{eqnarray}}

\newcommand{\threev}[3]{\left(\begin{tabular}{c} $#1$ \\ $#2$ \\ $#3$ \end{tabular}\right)}

\newcommand{\f}{\varphi}
\newcommand{\h}{{1\over2}}

\newcommand{\ind}[1]{\index{#1}#1}

\documentclass[a4paper]{article}
\usepackage{epsfig}
\usepackage{axodraw}
\usepackage{graphicx}
\usepackage{amsmath}
\usepackage{makeidx}
\usepackage[top=1in, bottom=1in, left=1in, right=1in]{geometry}

\begin{document}

\begin{center}
\LARGE
Path Integrals in Polar Field Variables in QFT\\[30pt]
\large
E.N. Argyres \hspace{50pt} C.G. Papadopoulos\footnote{Costas.Papadopoulos@cern.ch}\\[5pt]
Institute of Nuclear Physics, NCSR "Demokritos", Athens, Greece\\[20pt]
M.T.M. van Kessel\footnote{M.vanKessel@science.ru.nl} \hspace{50pt} R.H.P. Kleiss\footnote{R.Kleiss@science.ru.nl}\\[5pt]
IMAPP, FNWI, Radboud Universiteit Nijmegen, Nijmegen, The Netherlands\\[20pt]
October 21, 2008\\[50pt]
\end{center}

\section{Abstract}

We show how to transform a $d$-dimensional Euclidean path integral in terms of two (Cartesian) fields to a path integral in terms of polar field variables. First we present a conjecture that states how this transformation should be done. Then we show that this conjecture is correct in the case of two toy models. Finally the conjecture will be proven for a general QFT model with two fields.

\newpage

\section{Introduction}

Since Feynman presented his path-integral formulation of quantum mechanics \cite{Feynman} there have been a lot of applications for this formalism, both in quantum mechanics and in quantum field theory. In the case of quantum mechanics it is known how to transform a path integral in terms of the normal (Cartesian) fields to a path integral in terms of curvilinear fields. This is very convenient in calculations for models that possess a rotational symmetry. Such transformations are worked in the case of quantum mechanics by Edwards et al.\ \cite{Edwards}, Peak et al.\ \cite{Peak}, Lee \cite{Lee}, B\"ohm et al.\ \cite{Bohm}, Gerry et al.\ \cite{Gerry}, Grosche et al.\ \cite{Grosche} and Kleinert \cite{Kleinert}.

Quantum mechanics corresponds mathematically to one-dimensional quantum field theory. This means that the path integral in terms of curvilinear fields is known for quantum field theories in one dimension. For dimensions greater than one it is, up to now, not known how the path integral can be transformed into a path integral in terms of curvilinear field variables.

In this paper we will show how this transformation can be performed, in the case of a $d$-dimensional Euclidean quantum field theory with \emph{two} scalar fields.

The outline of this paper is as follows. First we shall discuss the difficulties that one encounters when trying to transform to polar field variables in a $d$-dimensional path integral. Then we present a conjecture on how this transformation should be done. We will verify this conjecture for two $d$-dimensional toy models. After this we will actually prove the conjecture. Finally we will specify our calculations to the one-dimensional case, to make contact with the literature on one-dimensional path integrals in terms of polar fields.

\section{The Path Integral}

The generic form of a $d$-dimensional Euclidean path integral $P$ in two scalar fields $\f_1$ and $\f_2$ is:
\bqa
& & P = \int \mathcal{D}\f_1\mathcal{D}\f_2 \; \f_1(x^{(1)})\cdots\f_1(x^{(m)}) \; \f_2(y^{(1)})\cdots\f_2(y^{(n)}) \cdot \nonumber\\
& & \phantom{P = \int \mathcal{D}\f_1\mathcal{D}\f_2 \;} \exp\left(-{1\over\hbar}\int d^dx\left( \h\left(\nabla\f_1\right)^2 + \h\left(\nabla\f_2\right)^2 + V(\f_1,\f_2) \right)\right) \label{genericpathint}
\eqa
Here $x^{(i)}$ and $y^{(i)}$ are $d$-dimensional vectors,
\bq
x \equiv \threev{x_1}{\vdots}{x_d} \;,
\eq
and $\nabla$ is the $d$-dimensional vector
\bq
\nabla \equiv \threev{\partial/\partial x_1}{\vdots}{\partial/\partial x_d} \;.
\eq

In (\ref{genericpathint}) we have written the path integral $P$ in continuum form. We have to realize that this is just a shorthand notation, and the path integral is only defined on the lattice. Therefore we should actually read (\ref{genericpathint}) as:
\bqa
P &=& M(\Delta,N) \int_{-\infty}^\infty \left( \prod_{i_1,\ldots,i_d=0}^{N-1} d\f_{1 \; i_1,\ldots,i_d} \right) \int_{-\infty}^\infty \left( \prod_{i_1,\ldots,i_d=0}^{N-1} d\f_{2 \; i_1,\ldots,i_d} \right) \cdot \nonumber\\[5pt]
& & \hspace{70pt} \f_{1 \; j^{(1)}_1,\ldots,j^{(1)}_d} \cdots \f_{1 \; j^{(m)}_1,\ldots,j^{(m)}_d} \; \f_{2 \; k^{(1)}_1,\ldots,k^{(1)}_d} \cdots \f_{2 \; k^{(n)}_1,\ldots,k^{(n)}_d} \cdot \nonumber\\[5pt]
& & \hspace{70pt} \exp\left(-{1\over\hbar}S\right) \label{genericpathintdiscr}
\eqa
Here we have introduced a $d$-dimensional lattice with $N$ points in each direction. We denote the lattice spacing by $\Delta$. The length of the lattice is $(N-1)\Delta \equiv L$ in each direction. The discrete coordinates $i$ are related to the continuum coordinates $x$ as:
\bq
x \equiv \threev{-L/2+\Delta i_1}{\vdots}{-L/2+\Delta i_d} \;, \quad i_1,\ldots,i_d=0,\ldots,N-1
\eq
Here we have chosen a specific correspondence between the continuum and discrete coordinates. For details on this correspondence see \cite{Lee,Grosche2}. Which correspondence we pick does not matter as long as the potential $V$ contains no products of derivatives of the fields and the fields themselves. This is what we will assume in the rest of this paper. The fields on the discrete lattice are denoted by:
\bqa
\f_{1 \;\; i_1,\ldots,i_d} &\equiv& \f_1(x) \nonumber\\
\f_{2 \;\; i_1,\ldots,i_d} &\equiv& \f_2(x)
\eqa
For these fields we shall assume periodic boundary conditions in all directions. For the 1-direction this means:
\bqa
\f_{1 \;\; N,i_2,\ldots,i_d} &=& \f_{1 \;\; 0,i_2,\ldots,i_d} \nonumber\\
\f_{2 \;\; N,i_2,\ldots,i_d} &=& \f_{2 \;\; 0,i_2,\ldots,i_d}
\eqa
The discrete coordinates $j^{(1)}$ up to $j^{(m)}$ and $k^{(1)}$ up to $k^{(n)}$ correspond respectively to the continuum coordinates $x^{(1)}$ up to $x^{(m)}$ and $y^{(1)}$ up to $y^{(n)}$. The action $S$ in (\ref{genericpathintdiscr}) is given by:
\bqa
S &=& \Delta^d \sum_{i_1,\ldots,i_d=0}^{N-1} \Bigg( \h{1\over\Delta^2}\left(\f_{1 \;\; i_1+1,i_2,\ldots,i_d}-\f_{1 \;\; i_1,i_2,\ldots,i_d}\right)^2 + \ldots + \nonumber\\
& & \phantom{\Delta^d \sum_{i_1,\ldots,i_d=0}^{N-1} \Bigg(} \h{1\over\Delta^2}\left(\f_{1 \;\; i_1,\ldots,i_{d-1},i_d+1}-\f_{1 \;\; i_1,\ldots,i_{d-1},i_d}\right)^2 + \nonumber\\
& & \phantom{\Delta^d \sum_{i_1,\ldots,i_d=0}^{N-1} \Bigg(} \h{1\over\Delta^2}\left(\f_{2 \;\; i_1+1,i_2,\ldots,i_d}-\f_{2 \;\; i_1,i_2,\ldots,i_d}\right)^2 + \ldots + \nonumber\\
& & \phantom{\Delta^d \sum_{i_1,\ldots,i_d=0}^{N-1} \Bigg(} \h{1\over\Delta^2}\left(\f_{2 \;\; i_1,\ldots,i_{d-1},i_d+1}-\f_{2 \;\; i_1,\ldots,i_{d-1},i_d}\right)^2 + \nonumber\\
& & \phantom{\Delta^d \sum_{i_1,\ldots,i_d=0}^{N-1} \Bigg(} V\left(\f_{1 \;\; i_1,\ldots,i_d},\f_{2 \;\; i_1,\ldots,i_d}\right) \Bigg) \label{discraction}
\eqa
The function $M$ in (\ref{genericpathintdiscr}) is a measure, which is quite unimportant for our purposes. In QFT we are never interested in the path integral itself, but rather in the ratio of two path integrals. In this ratio the measure $M$ always drops out. Details about this measure can be found in \cite{Lee,Grosche2}.

\section{Polar Field Variables}

When dealing with a normal integral a common technique to solve it is to transform to a different integration variable. Since the path integral (\ref{genericpathintdiscr}) is merely a $2n^d$-dimensional integral, one should also be able to transform to different integration variables in this case. In particular one can transform to polar field variables, $r$ and $\theta$, defined by:
\bqa
\f_1(x) &=& r(x)\cos\theta(x) \nonumber\\
\f_2(x) &=& r(x)\sin\theta(x) \;.
\eqa
It is this transformation to polar field variables that we shall study in this paper. We wish to emphasize here that \emph{not} the space-time coordinates, but the quantum fields $\f_1$ and $\f_2$ are transformed to polar fields $r$ and $\theta$.

The discreteness of the path integral makes such a variable transformation very complicated. Several difficult questions immediately pop up:
\begin{enumerate}
\item In principle the whole path integral is only defined on a lattice, so the transformation should also be done with the path integral in discrete form. This means one \emph{cannot} simply let the transformation work on the continuum action. Instead one has to write out the action in discrete form, as in (\ref{discraction}), and only then let the transformation work. After this one gets a complicated action, with also terms proportional to the lattice spacing $\Delta$. These terms \emph{cannot} be discarded, since one has to perform the path integration first and then take the continuum limit $\Delta\rightarrow\infty$. It is not obvious that the terms proportional to $\Delta$ will not give a finite contribution in this continuum limit. In fact, we will see that they \emph{do} give finite contributions to Green's functions.
\item After the transformation the domain of integration is not $\langle-\infty,\infty\rangle$. For the $r$-variables it is $[0,\infty\rangle$, whereas for the $\theta$-variables it is $[-\pi,\pi]$. How does one evaluate such a path integral, especially because we can only compute path integrals with perturbation theory? To do perturbation theory we have to be able to identify a Gaussian part, and the fields in such a Gaussian part are always integrated from $-\infty$ to $\infty$.
\item After the transformation one gets a Jacobian, how does one deal with this? Since we can only do perturbation theory we also have to identify the Gaussian part and the perturbative part of this Jacobian.
\end{enumerate}

From these questions it is clear that transforming to polar variables in a $d$-dimensional path integral is very complicated. For one-dimensional systems, i.e.\ quantum-mechanical systems, there is quite some literature on the transformation to polar variables.

In his textbook \cite{Lee} Lee derives the quantum-mechanical path integral in curvilinear coordinates in chapter 19. The result (19.49) is a path integral with a new action $L_{\textrm{eff}}$, which is \emph{not} equal to the action one would find by transforming to polar coordinates in the \emph{continuum} action (in Cartesian coordinates).

Edwards et al.\ \cite{Edwards} and Peak et al.\ \cite{Peak} also transform to polar coordinates in the discrete quantum-mechanical path integral. They find that terms of order $\Delta$ or higher, which arise when transforming to polar coordinates in the discrete action (in Cartesian coordinates), \emph{cannot} all be neglected.

\section{A Conjecture}

From the above it should be clear that transforming a path integral in terms of the normal (i.e.\ Cartesian) fields $\f_1$ and $\f_2$ to a path integral in terms of polar fields is far from trivial. To be able to do computations at all we will present a conjecture in this section. In the next sections we will then try to make this conjecture plausible by considering certain toy models where we can see that the conjecture actually works.

The generic form of a $d$-dimensional Euclidean path integral $P$ in two fields is given in (\ref{genericpathint}). Very naively, one could hope that the transformation to polar variables works as:
\bqa
\int_{-\infty}^\infty \mathcal{D}\f_1 \int_{-\infty}^\infty \mathcal{D}\f_2 &\rightarrow& \int_{-\infty}^\infty \mathcal{D}r \; \prod_x r(x) \int_{-\infty}^\infty \; \mathcal{D}\theta \nonumber\\
\f_1(x) &\rightarrow& r(x)\cos\theta(x) \nonumber\\
\f_2(x) &\rightarrow& r(x)\sin\theta(x) \nonumber\\
\int d^dx\left( \h\left(\nabla\f_1\right)^2 + \h\left(\nabla\f_2\right)^2 \right) &\rightarrow& \int d^dx\left( \h\left(\nabla r\right)^2 + \h r^2\left(\nabla\theta\right)^2 \right) \nonumber\\
\int d^dx \;  V(\f_1,\f_2) &\rightarrow& \int d^dx \; V(r\cos\theta,r\sin\theta)
\eqa
Here one has just extended the integration domains for $r$ and $\theta$ to $\langle-\infty,\infty\rangle$. In the fourth line one has just transformed the continuum action to an action in terms of polar fields, disregarding the fact that one should do this on the lattice, where the path integral is defined.

To bring all these expressions in continuum form we still have to do something about the Jacobian factor
\bq
\prod_x r(x) \;,
\eq
since this is still a lattice expression. We shall write this Jacobian as:
\bqa
\prod_x r(x) &=& \prod_x \exp\left(-{1\over\hbar}\left(-\hbar\ln r(x)\right)\right) \nonumber\\
&=& \exp\left(-{1\over\hbar} \sum_x \left(-\hbar\ln r(x)\right)\right) \nonumber\\
&\rightarrow& \exp\left(-{1\over\hbar}{1\over\Delta^d} \int d^dx \left(-\hbar\ln r(x)\right)\right) \nonumber\\
&\rightarrow& \exp\left(-{1\over\hbar} \left[{1\over(2\pi)^d}\int d^dk\right] \int d^dx \left(-\hbar\ln r(x)\right)\right)
\eqa
In a discrete space (i.e.\ finite $\Delta$) it can easily be seen that the last step is correct. Now we have also written the Jacobian as a continuum expression, which looks a lot like the exponential of an action. 

So, we might hope that the continuum form of a path integral in polar field variables is given by making the substitutions
\bqa
\int_{-\infty}^\infty \mathcal{D}\f_1 \int_{-\infty}^\infty \mathcal{D}\f_2 &\rightarrow& \int_{-\infty}^\infty \mathcal{D}r \; \int_{-\infty}^\infty \; \mathcal{D}\theta \cdot \nonumber\\
& & \quad \exp\left(-{1\over\hbar} \left[{1\over(2\pi)^d}\int d^dk\right] \int d^dx \left(-\hbar\ln r(x)\right)\right) \nonumber\\
\f_1(x) &\rightarrow& r(x)\cos\theta(x) \nonumber\\
\f_2(x) &\rightarrow& r(x)\sin\theta(x) \nonumber\\
\int d^dx\left( \h\left(\nabla\f_1\right)^2 + \h\left(\nabla\f_2\right)^2 \right) &\rightarrow& \int d^dx\left( \h\left(\nabla r\right)^2 + \h r^2\left(\nabla\theta\right)^2 \right) \nonumber\\
\int d^dx \;  V(\f_1,\f_2) &\rightarrow& \int d^dx \; V(r\cos\theta,r\sin\theta) \label{conjecture}
\eqa
in the path integral in Cartesian form.

Now \ind{the conjecture} we are going to make is:\\[15pt]
\fbox{\parbox{15.64cm}{\textbf{Conjecture} \;\; \emph{It is correct to transform to polar variables naively, as in (\ref{conjecture}), provided one does the calculation in a $d$-dimensional way.}}}\\[15pt]
This seems a very strange conjecture, and with all the remarks we made it is hard to imagine how it can work. However in the next two sections we shall demonstrate that indeed this conjecture is true for two toy models. Also there we will demonstrate what is meant exactly by `calculating in a $d$-dimensional way'.

\section{The Shifted Toy Model}

In this section we shall calculate several Green's functions in the so-called \ind{shifted toy model}. This model has an action
\bq
S = \int d^dx\left( \h\left(\nabla\f_1\right)^2 + \h\left(\nabla\f_2\right)^2 + \h m^2(\f_1-v)^2 + \h m^2\f_2^2 \right) \;.
\eq
This is just the action of a free model with the $\f_1$-field shifted, hence the name. We shifted this field such that the minimum of the action is at $\f_1=v,\;\f_2=0$, and \emph{not} at $r=0$, because the transformation to polar fields becomes singular at this point.

To prove that the conjecture indeed works in the case of this model we shall now calculate several Green's functions through the normal, Cartesian, path integral \emph{and} through the path integral in terms of polar fields. Then we can compare results.

\subsection{Cartesian Results}

Because the shifted toy model is just a free theory with one field shifted it is very easy to obtain the \emph{exact full} Green's functions. They are:
\bqa
\langle \f_1(x) \rangle &=& v \nonumber\\
\langle \f_2(x) \rangle &=& 0 \nonumber\\
\langle \f_1(x) \f_1(y) \rangle &=& v^2 + {\hbar\over(2\pi)^d} \int d^dk \; \frac{e^{ik \cdot (x-y)}}{k^2+m^2} \nonumber\\
&=& v^2 + \hbar \; A_m(x-y) \nonumber\\
\langle \f_2(x) \f_2(y) \rangle &=& {\hbar\over(2\pi)^d} \int d^dk \; \frac{e^{ik \cdot (x-y)}}{k^2+m^2} \nonumber\\
&=& \hbar \; A_m(x-y) \nonumber\\
\langle \f_1(x) \f_2(y) \rangle &=& 0
\eqa
Here we wrote the results in terms of standard $d$-dimensional integrals, which are defined in appendix \ref{appstandint}.

\subsection{Polar Results}

Now we perform the transformation to polar fields in the path integral:
\bqa
\f_1(x) &=& r(x)\cos\left({w(x)\over v}\right) \nonumber\\
\f_2(x) &=& r(x)\sin\left({w(x)\over v}\right)
\eqa
Here we have used $w(x)/v$ instead of $\theta(x)$ to have a new angular field-variable with also the dimensions of a field. This is purely a matter of convenience.

To calculate the Green's functions through the path integral in polar fields now we will use the conjecture. According to this conjecture, the new action we have to work with is:
\bq
S = \int d^dx \left( {1\over2} \left( \nabla r(x) \right)^2 + {1\over2v^2} r^2(x) \left( \nabla w(x) \right)^2 + {1\over2}m^2 r^2(x) - m^2v r(x) \cos(w(x)/v) \right).
\eq
The minimum of the action is at $r(x)=v$ and $w(x)=0$. Expanding around $r(x)=v$ and writing
\bq
r(x) \equiv v + \eta(x)
\eq
we get
\bqa
S &=& \int d^dx \bigg( {1\over2} \left( \nabla \eta(x) \right)^2 + {1\over2} \left( \nabla w(x) \right)^2 + {1\over v} \eta(x) \left( \nabla w(x) \right)^2 + {1\over2v^2} \eta^2(x) \left( \nabla w(x) \right)^2 + \nonumber\\
& & \phantom{\int d^dx \bigg(} m^2v \eta(x) + {1\over2}m^2 \eta^2(x) - m^2v^2 \cos(w(x)/v) - m^2v \eta(x) \cos(w(x)/v) \bigg) \;. \nonumber\\
\eqa
Expanding also around $w(x)=0$, i.e.\ expanding the cosine gives:
\bqa
S &=& \int d^dx \bigg( {1\over2} \left( \nabla \eta(x) \right)^2 + {1\over2} \left( \nabla w(x) \right)^2 + {1\over v} \eta(x) \left( \nabla w(x) \right)^2 + {1\over2v^2} \eta^2(x) \left( \nabla w(x) \right)^2 + \nonumber\\
& & \phantom{\int d^dx \bigg(} {1\over2}m^2 \eta^2(x) + {1\over2}m^2 w^2(x) - {m^2\over24v^2} w^4(x) + {m^2\over2v} \eta(x) w^2(x) - {m^2\over24v^3} \eta(x) w^4(x) + \nonumber\\
& & \phantom{\int d^dx \bigg(} \mathcal{O}(\hbar^3) \bigg).
\eqa

The Jacobian gives, according to the conjecture:
\bqa
& & \exp\left(-{1\over\hbar} I \int d^dx \left(-\hbar\ln r(x)\right)\right) \sim \nonumber\\
& & \exp\left(-{1\over\hbar} I \int d^dx \left(-\hbar\ln\left(1+{\eta(x)\over v}\right)\right)\right) = \nonumber\\
& & \exp\left(-{1\over\hbar} I \int d^dx \left(-\hbar \sum_{n=1}^\infty \frac{(-1)^{n+1}}{n} \left({\eta(x)\over v}\right)^n \right)\right)
\eqa
The standard integral $I$ is defined in appendix \ref{appstandint}.

Now can read off the Feynman rules from the action and the Jacobian, and the conjecture states that we can calculate everything in the continuum, provided we do a $d$-dimensional calculation. The Feynman rules are (up to order $\hbar^{5/2}$):
{\allowdisplaybreaks\bqa
\begin{picture}(100, 20)(0, 17)
\Line(20, 20)(80, 20)
\end{picture}
&\leftrightarrow& \frac{\hbar}{k^2+m^2} \nonumber\\
\begin{picture}(100, 20)(0, 17)
\DashLine(20, 20)(80, 20){5}
\end{picture}
&\leftrightarrow& \frac{\hbar}{k^2+m^2} \nonumber\\
\begin{picture}(100, 40)(0, 17)
\Line(20, 20)(50, 20)
\DashLine(50, 20)(80, 0){5}
\DashLine(50, 20)(80, 40){5}
\Text(80,9)[c]{$k_1$}
\Text(80,31)[c]{$k_2$}
\end{picture}
&\leftrightarrow& {2\over\hbar v} k_1 \cdot k_2 - {m^2\over\hbar v} \nonumber\\
\begin{picture}(100, 40)(0, 17)
\Line(20,40)(50,20)
\Line(20,0)(50,20)
\DashLine(80,0)(50,20){5}
\DashLine(80,40)(50,20){5}
\Text(80,9)[c]{$k_1$}
\Text(80,31)[c]{$k_2$}
\end{picture}
&\leftrightarrow& {2\over\hbar v^2} k_1 \cdot k_2 \nonumber\\
\begin{picture}(100, 50)(0, 17)
\DashLine(20, 0)(80, 40){5}
\DashLine(20, 40)(80, 0){5}
\end{picture}
&\leftrightarrow& {m^2\over\hbar v^2} \nonumber\\
\begin{picture}(100, 40)(0, 17)
\Line(20, 20)(50, 20)
\DashLine(20, 0)(80, 40){5}
\DashLine(20, 40)(80, 0){5}
\end{picture}
&\leftrightarrow& {m^2\over\hbar v^3} \nonumber\\
\begin{picture}(100, 40)(0, 17)
\Line(20, 20)(50, 20)
\Vertex(50, 20){3}
\end{picture}
&\leftrightarrow& {1\over v} \; I \nonumber\\
\begin{picture}(100, 40)(0, 17)
\Line(20, 20)(80, 20)
\Vertex(50, 20){3}
\end{picture}
&\leftrightarrow& -{1\over v^2} \; I \nonumber\\
\begin{picture}(100, 40)(0, 17)
\Line(20, 20)(50, 20)
\Line(50, 20)(80, 40)
\Line(50, 20)(80, 0)
\Vertex(50, 20){3}
\end{picture}
&\leftrightarrow& {2\over v^3} \; I
\eqa}\\[5pt]
Here all momenta are counted as going into the vertex. With these rules we can now compute Green's functions up to order $\hbar^2$.

First we will demonstrate however what we mean exactly by `calculating in a $d$-dimensional way'. What we mean can most easily be seen in the following `$d$-dimensional calculation' of a tadpole diagram.
\vspace{-15pt}
{\allowdisplaybreaks\bqa
\begin{picture}(100,50)(0,25)
\Line(0,25)(50,25)
\DashCArc(75,25)(25,0,360){3}
\Line(75,0)(75,50)
\end{picture} &=& \h{\hbar\over m^2} {1\over(2\pi)^{2d}}\int d^dk \; d^dl \; \left(-{2\over\hbar v}k^2-{m^2\over\hbar v}\right) \left(-{2\over\hbar v}k\cdot l-{m^2\over\hbar v}\right)^2 \nonumber\\
& & \phantom{\h{\hbar\over m^2} {1\over(2\pi)^{2d}}\int d^dk \; d^dl \;} \left(\frac{\hbar}{k^2+m^2}\right)^2 \frac{\hbar}{l^2+m^2} \frac{\hbar}{(k-l)^2+m^2} \nonumber\\
&=& \h{\hbar^2\over m^2v^3} {1\over(2\pi)^{2d}}\int d^dk \; d^dl \; (-2k^2-m^2)(-2k\cdot l-m^2)^2 \nonumber\\
& & \phantom{\h{\hbar^2\over m^2v^3} {1\over(2\pi)^{2d}}\int d^dk \; d^dl \;} \left(\frac{1}{k^2+m^2}\right)^2 \frac{1}{l^2+m^2} \frac{1}{(k-l)^2+m^2} \nonumber\\
&=& -{\hbar^2\over m^2v^3} {1\over(2\pi)^{2d}}\int d^dk \; d^dl \; (-2k\cdot l-m^2)^2 \cdot \nonumber\\
& & \phantom{-{\hbar^2\over m^2v^3} {1\over(2\pi)^{2d}}\int d^dk \; d^dl \;} \frac{1}{k^2+m^2} \frac{1}{l^2+m^2} \frac{1}{(k-l)^2+m^2} + \nonumber\\
& & \h{\hbar^2\over v^3} {1\over(2\pi)^{2d}}\int d^dk \; d^dl \; (-2k\cdot l-m^2)^2 \cdot \nonumber\\
& & \phantom{\h{\hbar^2\over v^3} {1\over(2\pi)^{2d}}\int d^dk \; d^dl \;} \left(\frac{1}{k^2+m^2}\right)^2 \frac{1}{l^2+m^2} \frac{1}{(k-l)^2+m^2} \nonumber\\
\eqa}
Now use
\bq
-2k\cdot l-m^2 = \left[(k-l)^2+m^2\right] - \left[k^2+m^2\right] - \left[l^2+m^2\right] \;.
\eq
Then the tadpole diagram becomes:
\bqa
& & -{\hbar^2\over m^2v^3} {1\over(2\pi)^{2d}}\int d^dk \; d^dl \; (-2k\cdot l-m^2) \nonumber\\
& & \phantom{-{\hbar^2\over m^2v^3} {1\over(2\pi)^{2d}}\int d^dk \; d^dl \;} \left( \frac{1}{k^2+m^2} \frac{1}{l^2+m^2} - 2\frac{1}{k^2+m^2} \frac{1}{(k-l)^2+m^2} \right) + \nonumber\\
& & \h{\hbar^2\over v^3} {1\over(2\pi)^{2d}}\int d^dk \; d^dl \; (-2k\cdot l-m^2) \nonumber\\
& & \phantom{\h{\hbar^2\over v^3} {1\over(2\pi)^{2d}}\int d^dk \; d^dl \;} \bigg( \frac{1}{(k^2+m^2)^2} \frac{1}{l^2+m^2} - \frac{1}{k^2+m^2} \frac{1}{l^2+m^2} \frac{1}{(k-l)^2+m^2} + \nonumber\\
& & \phantom{\h{\hbar^2\over v^3} {1\over(2\pi)^{2d}}\int d^dk \; d^dl \; \bigg(} -\frac{1}{(k^2+m^2)^2} \frac{1}{(k-l)^2+m^2} \bigg) \nonumber\\
&=& {9\over2}{\hbar^2\over v^3} \; I(0,m)^2 - 4{\hbar^2\over m^2v^3} \; I(0,m) I - {\hbar^2m^2\over v^3} \; I(0,m)I(0,m,0,m)
\eqa
In the above steps, and in all $d$-dimensional calculations, one essentially uses three rules: One writes dot-products from the vertices in terms of the denominators of the propagators, to let them cancel as much as possible, one can shift all loop momenta and one can set
\bq
\int d^dk \; \frac{k_i}{k^2+m^2} = 0 \;.
\eq
Using these three rules is what we mean by a `$d$-dimensional calculation'.

Notice that, for example in dimension one, where $k\cdot l$ becomes a simple product, we could also have combined momenta coming from different vertices and let them cancel denominators. This simplifies the calculation of the tadpole diagram considerably, however this is \emph{not} what we mean by a `$d$-dimensional calculation'. The result in terms of standard integrals is also \emph{different}. Even the numerical result is \emph{different} because the diagram contains a divergence, also in $d=1$. (The divergence comes from $I$.) In order for the conjecture to work we \emph{have} to perform a $d$-dimensional calculation.

Now we compute $\langle\f_1(x)\rangle$, $\langle\f_2(x)\rangle$, $\langle\f_1(x)\f_1(y)\rangle$, $\langle\f_2(x)\f_2(y)\rangle$ and $\langle\f_1(x)\f_2(y)\rangle$ via the polar fields, up to order $\hbar^2$. To do this we first have to express these Green's functions in terms of the $\eta$- and $w$-Green's-functions.
{\allowdisplaybreaks\bqa
\langle\f_1(x)\rangle &=& \langle(v+\eta(x))\cos(w(x)/v)\rangle \nonumber\\
&=& v + \langle\eta(x)\rangle - {1\over2v}\langle w^2(x)\rangle - {1\over2v^2}\langle\eta(x)w^2(x)\rangle + {1\over24v^3}\langle w^4(x)\rangle + \mathcal{O}(\hbar^3) \nonumber\\
\langle\f_2(x)\rangle &=& \langle(v+\eta(x))\sin(w(x)/v)\rangle \nonumber\\
&=& \langle w(x)\rangle + {1\over v}\langle\eta(x)w(x)\rangle - {1\over6v^2}\langle w^3(x)\rangle - {1\over6v^3}\langle\eta(x)w^3(x)\rangle + \mathcal{O}(\hbar^3) \nonumber\\
\langle\f_1(x)\f_1(y)\rangle &=& \langle(v+\eta(x))(v+\eta(y))\cos(w(x)/v)\cos(w(y)/v)\rangle \nonumber\\
&=& v^2 + 2v\langle\eta(x)\rangle + \langle\eta(x)\eta(y)\rangle - \langle w^2(x)\rangle - {1\over v}\langle\eta(x)w^2(x)\rangle + \nonumber\\
& & -{1\over2v}\langle\eta(x)w^2(y)\rangle - {1\over2v}\langle\eta(y)w^2(x)\rangle + {1\over12v^2}\langle w^4(x)\rangle + \nonumber\\
& & {1\over4v^2}\langle w^2(x)w^2(y)\rangle - {1\over2v^2}\langle\eta(x)\eta(y)w^2(x)\rangle - {1\over2v^2}\langle\eta(x)\eta(y)w^2(y)\rangle + \nonumber\\[5pt]
& & \mathcal{O}(\hbar^3) \nonumber\\[5pt]
\langle\f_2(x)\f_2(y)\rangle &=& \langle(v+\eta(x))(v+\eta(y))\sin(w(x)/v)\sin(w(y)/v)\rangle \nonumber\\
&=& \langle w(x)w(y)\rangle + {1\over v}\langle\eta(x)w(x)w(y)\rangle + {1\over v}\langle\eta(y)w(x)w(y)\rangle + \nonumber\\
& & {1\over v^2}\langle\eta(x)\eta(y)w(x)w(y)\rangle - {1\over6v^2}\langle w^3(x)w(y)\rangle - {1\over6v^2}\langle w(x)w^3(y)\rangle + \nonumber\\[5pt]
& & \mathcal{O}(\hbar^3) \nonumber\\[5pt]
\langle\f_1(x)\f_2(y)\rangle &=& \langle(v+\eta(x))(v+\eta(y))\cos(w(x)/v)\sin(w(y)/v)\rangle \nonumber\\
&=& v\langle w(y)\rangle + \langle\eta(x)w(y)\rangle + \langle\eta(y)w(y)\rangle - {1\over6v}\langle w^3(y)\rangle + \nonumber\\
& & -{1\over2v}\langle w^2(x)w(y)\rangle + {1\over v}\langle\eta(x)\eta(y)w(y)\rangle - {1\over6v^2}\langle\eta(x)w^3(y)\rangle + \nonumber\\
& & -{1\over6v^2}\langle\eta(y)w^3(y)\rangle - {1\over2v^2}\langle\eta(x)w^2(x)w(y)\rangle - {1\over2v^2}\langle\eta(y)w^2(x)w(y)\rangle + \nonumber\\[5pt]
& & \mathcal{O}(\hbar^3) \label{Cartaverages}
\eqa}
Notice that in these formulas all Green's-functions are \emph{full} Green's functions, i.e.\ all Green's functions contain the connected \emph{and} disconnected part. We wish to stress here that we \emph{cannot} replace all averages $\langle\ldots\rangle$ by connected averages $\langle\ldots\rangle_c$ on both sides of the equality sign in (\ref{Cartaverages}). Connected does \emph{not} mean the same in the Cartesian- and the polar-field formalism. In other words taking the connected part and transforming to polar fields are two operations that do \emph{not} commute.

Now we compute all the $\eta$- and $w$-Green's-functions that we need, up to order $\hbar^2$. All results will be expressed in the standard integrals from appendix \ref{appstandint}. From the Feynman rules we can immediately see that any Green's function with an odd number of $w$'s is zero.

\subsubsection{The $\eta$-Tadpole}

Below we list all the diagrams contributing to $\langle\tilde{\eta}\rangle=\langle\eta\rangle$ up to order $\hbar^2$.
\vspace{-25pt}
{\allowdisplaybreaks\bqa
\begin{picture}(100,50)(0,25)
\Line(0,25)(50,25)
\DashCArc(75,25)(25,0,360){3}
\end{picture} \qquad &=& \qquad -{\hbar\over m^2v} \; I + {1\over2}{\hbar\over v} \; I(0,m) \nonumber\\
\begin{picture}(100,50)(0,25)
\Line(0,25)(50,25)
\Vertex(50,25){4}
\end{picture} \qquad &=& \qquad {\hbar\over m^2v} \; I \nonumber\\
\begin{picture}(100,50)(0,25)
\Line(0,25)(25,25)
\GCirc(37.5,25){12.5}{0.5}
\Line(50,25)(75,25)
\GCirc(87.5,25){12.5}{0.5}
\end{picture} \qquad &=& \qquad -{1\over2}\frac{\hbar^2}{v^3} \; I(0,m)^2 + {1\over4}\frac{\hbar^2m^2}{v^3} \; I(0,m,0,m) \; I(0,m) \nonumber\\
\begin{picture}(100,50)(0,25)
\Line(0,25)(50,25)
\DashCArc(75,25)(25,0,360){3}
\Line(75,0)(75,50)
\end{picture} \qquad &=& \qquad -4{\hbar^2\over m^2v^3} \; I(0,m) \; I + {9\over2}\frac{\hbar^2}{v^3} \; I(0,m)^2 + \nonumber\\
& & \qquad -\frac{\hbar^2m^2}{v^3} \; I(0,m,0,m) \; I(0,m) \nonumber\\
\begin{picture}(100,50)(0,25)
\Line(0,25)(100,25)
\DashCArc(75,25)(25,0,360){3}
\end{picture} \qquad &=& \qquad 2{\hbar^2\over m^2v^3} \; I(0,m) \; I - 2\frac{\hbar^2}{v^3} \; I(0,m)^2 \nonumber\\
\begin{picture}(100,50)(0,25)
\Line(0,25)(20,25)
\DashCArc(40,25)(20,0,360){3}
\DashCArc(80,25)(20,0,360){3}
\end{picture} \qquad &=& \qquad -{1\over2}\frac{\hbar^2}{v^3} \; I(0,m)^2 + {1\over4}\frac{\hbar^2m^2}{v^3} \; I(0,m,0,m) \; I(0,m) \nonumber\\
\begin{picture}(100,50)(0,25)
\Line(0,25)(20,25)
\DashCArc(40,25)(20,0,360){3}
\CArc(80,25)(20,0,360)
\end{picture} \qquad &=& \qquad {\hbar^2\over m^2v^3} \; I(0,m) \; I - {3\over2}\frac{\hbar^2}{v^3} \; I(0,m)^2 + \nonumber\\
& & \qquad \h\frac{\hbar^2m^2}{v^3} \; I(0,m,0,m) \; I(0,m) \nonumber\\
\begin{picture}(100,50)(0,25)
\Line(0,25)(75,25)
\DashCArc(75,37.5)(12.5,0,360){3}
\DashCArc(75,12.5)(12.5,0,360){3}
\end{picture} \qquad &=& \qquad {1\over8} \frac{\hbar^2}{v^3} \; I(0,m)^2 \nonumber\\
\begin{picture}(100,50)(0,25)
\Line(0,25)(50,25)
\CArc(75,25)(25,0,360)
\Vertex(50,25){4}
\end{picture} \qquad &=& \qquad {\hbar^2\over m^2v^3} \; I(0,m) \; I
\eqa}\\[15pt]
The complete result for the $\eta$-tadpole is:
\bq
\langle\tilde{\eta}\rangle = \h{\hbar\over v} \; I(0,m) + {1\over8}{\hbar^2\over v^3} \; I(0,m)^2.
\eq
Notice that all the (badly divergent) $I$-integrals from the Jacobian have nicely cancelled against identical terms from $w$-loops.

\subsubsection{The $\eta$-Propagator}

Below we list the diagrams contributing to the \emph{connected} momentum-space $\eta$-propagator $\langle\tilde{\eta}(p)\tilde{\eta}(-p)\rangle_{\mathrm{c}}$.
\vspace{-18pt}
\bqa
\begin{picture}(70,40)(0,18)
\Line(0,20)(70,20)
\end{picture} \quad &=& \frac{\hbar}{p^2+m^2} \nonumber\\
\begin{picture}(70,40)(0,18)
\Line(0,20)(20,20)
\DashCArc(35,20)(15,-180,180){3}
\Line(50,20)(70,20)
\end{picture} \quad &=& 2{\hbar^2\over v^2} \frac{1}{(p^2+m^2)^2} \; I - {\hbar^2\over v^2}\frac{p^2+2m^2}{(p^2+m^2)^2} \; I(0,m) + \h{\hbar^2\over v^2}\;  I(0,m,p,m) \nonumber\\
\begin{picture}(70,40)(0,18)
\Line(0,5)(70,5)
\DashCArc(35,20)(15,-90,270){3}
\end{picture} \quad &=& -{\hbar^2\over v^2} \frac{1}{(p^2+m^2)^2} \; I + {\hbar^2m^2\over v^2}\frac{1}{(p^2+m^2)^2} \; I(0,m) \nonumber\\
\begin{picture}(70,40)(0,18)
\Line(0,20)(70,20)
\Vertex(35,20){4}
\end{picture} \quad &=& -{\hbar^2\over v^2} \frac{1}{(p^2+m^2)^2} \; I
\eqa\\
For the connected $\eta$-propagator we get:
\bq
\langle \tilde{\eta}(p)\tilde{\eta}(-p) \rangle_c = \frac{\hbar}{p^2+m^2} - {\hbar^2\over v^2}\frac{1}{p^2+m^2} \; I(0,m) + \h{\hbar^2\over v^2} I(0,m,p,m) 
\eq
Notice that also here the $I$-integrals cancel.

\subsubsection{The $w$-Propagator}

For the \emph{connected} $w$-propagator $\langle\tilde{w}(p)\tilde{w}(-p)\rangle_c$ we have the following contributions, up to order $\hbar^2$. 
\vspace{-17pt}
\bqa
\langle\tilde{w}(p)\tilde{w}(-p)\rangle_c &=&
\begin{picture}(70,40)(0,18)
\DashLine(0,20)(70,20){3}
\end{picture} \quad + \quad
\begin{picture}(70,40)(0,18)
\DashLine(0,20)(20,20){3}
\DashCArc(35,20)(15,-180,0){3}
\CArc(35,20)(15,0,180)
\DashLine(50,20)(70,20){3}
\end{picture} \quad + \quad
\begin{picture}(70,40)(0,18)
\DashLine(0,5)(70,5){3}
\Line(35,5)(35,25)
\GCirc(35,25){10}{0.5}
\end{picture} \quad + \nonumber\\
& & \begin{picture}(70,40)(0,18)
\DashLine(0,5)(70,5){3}
\DashCArc(35,20)(15,-90,270){3}
\end{picture} \quad + \quad 
\begin{picture}(70,40)(0,18)
\DashLine(0,5)(70,5){3}
\CArc(35,20)(15,-90,270)
\end{picture} \nonumber\\[20pt]
&=& \frac{\hbar}{p^2+m^2} + {\hbar^2\over v^2} \; I(0,m,p,m)
\eqa\\
Notice that in none of the diagrams in the connected $w$-propagator a (badly divergent) $I$-integral occurs.

The higher-point $\eta$- and $w$-Green's-functions up to order $\hbar^2$ are easy to calculate, for example $\langle\tilde{\eta}(p_1)\tilde{w}(p_2)\tilde{w}(p_3)\rangle_c$ is just one tree diagram.

Knowing the $\eta$- and $w$-Green's-functions in momentum space we can easily write down the configuration space Green's functions needed in (\ref{Cartaverages}), up to order $\hbar^2$. For details on this calculation we refer to one of the author's PhD-thesis \cite{vanKessel} (chapter 8). Finally substituting all the results in (\ref{Cartaverages}) gives us:
\bqa
\langle \f_1(x) \rangle &=& v \nonumber\\
\langle \f_2(x) \rangle &=& 0 \nonumber\\
\langle \f_1(x) \f_1(y) \rangle &=& v^2 + \hbar \; A_m(x-y) \nonumber\\
\langle \f_2(x) \f_2(y) \rangle &=& \hbar \; A_m(x-y) \nonumber\\
\langle \f_1(x) \f_2(y) \rangle &=& 0
\eqa
These are indeed the correct results for the Green's functions. So the conjecture is verified for several Green's functions in the shifted toy model up to order $\hbar^2$.

\section{The Arctangent Toy Model}

As another illustration of the conjecture we now consider the arctangent toy model. The action of this model is:
\bq \label{Cartactionarctan}
S = \int dx \left( \h(\nabla\f_1)^2 + \h(\nabla\f_2)^2 + \frac{m^2}{4v^2}(\f_1^2+\f_2^2-v^2)^2 + \frac{m^2v^4}{2(\f_1^2+\f_2^2)} \arctan^2 \left( \frac{\f_2}{\f_1} \right) \right)
\eq
This action has a single minimum at
\bq \label{minarctantoy}
\f_1 = v \;, \quad \f_2 = 0 \;.
\eq

We want to stress that this model is not at all a physical model. The action has an infinite number of vertices (by expanding the arctangent), which means this model is \emph{not} renormalizable. We just want to use this model as a toy model to test the conjecture. Especially because it is not renormalizable, so \emph{no} big cancellations can be expected to occur, the arctangent toy model is a very good test of the conjecture.

\subsection{Cartesian Results} \label{arctanCartsect}

To find the Cartesian Green's functions we expand the action around the minimum (\ref{minarctantoy}):
\bq
\f_1(x) = v + \eta_1(x) \;, \quad \f_2(x) = \eta_2(x) \;.
\eq
Notice that also the arctangent in the action has to be expanded, this term will give an infinite number of vertices. Up to order $\hbar^{5/2}$ the Feynman rules are:
{\allowdisplaybreaks\bqa
\begin{picture}(100, 20)(0, 17)
\Line(20, 20)(80, 20)
\end{picture}
&\leftrightarrow& \frac{\hbar}{k^2+\mu^2} \nonumber\\
\begin{picture}(100, 20)(0, 17)
\DashLine(20, 20)(80, 20){5}
\end{picture}
&\leftrightarrow& \frac{\hbar}{k^2+m^2} \nonumber\\
\begin{picture}(100, 40)(0, 17)
\Line(20, 20)(50, 20)
\Line(50, 20)(80, 0)
\Line(50, 20)(80, 40)
\end{picture}
&\leftrightarrow& -\frac{1}{\hbar}\frac{6m^2}{v} \nonumber\\
\begin{picture}(100, 40)(0, 17)
\Line(20, 20)(50, 20)
\DashLine(50, 20)(80, 0){5}
\DashLine(50, 20)(80, 40){5}
\end{picture}
&\leftrightarrow& \frac{1}{\hbar}\frac{2m^2}{v} \nonumber\\
\begin{picture}(100, 40)(0, 17)
\Line(20, 0)(80, 40)
\Line(20, 40)(80, 0)
\end{picture}
&\leftrightarrow& -\frac{1}{\hbar}\frac{6m^2}{v^2} \nonumber\\
\begin{picture}(100, 40)(0, 17)
\DashLine(20, 0)(80, 40){5}
\Line(20, 40)(80, 0)
\end{picture}
&\leftrightarrow& -\frac{1}{\hbar}\frac{22m^2}{v^2} \nonumber\\
\begin{picture}(100, 50)(0, 17)
\DashLine(20, 0)(80, 40){5}
\DashLine(20, 40)(80, 0){5}
\end{picture}
&\leftrightarrow& \frac{1}{\hbar}\frac{14m^2}{v^2} \nonumber\\
\begin{picture}(100, 40)(0, 17)
\Line(20, 0)(50, 20)
\Line(20, 20)(50, 20)
\Line(20, 40)(50, 20)
\DashLine(50, 20)(80, 40){5}
\DashLine(50, 20)(80, 0){5}
\end{picture}
&\leftrightarrow& \frac{1}{\hbar}\frac{120m^2}{v^3} \nonumber\\
\begin{picture}(100, 40)(0, 17)
\Line(20, 20)(50, 20)
\DashLine(20, 0)(80, 40){5}
\DashLine(20, 40)(80, 0){5}
\end{picture}
&\leftrightarrow& -\frac{1}{\hbar}\frac{120m^2}{v^3}
\eqa}\\[10pt]
Here the solid lines indicate the $\eta_1$-particle, the dashed lines indicate the $\eta_2$-particle and $\mu$ is given by $\mu = \sqrt{2}m$. With these Feynman rules we can now compute some Green's functions up to order $\hbar^2$. We shall not present all diagrams here, since this Cartesian calculation is straightforward and quite lengthy.
{\allowdisplaybreaks\bqa
\langle\f_1(x)\rangle &=& v + \langle\eta_1(x)\rangle \nonumber\\
&=& v - {3\over2}{\hbar\over v} \; I(0,\mu) + \h{\hbar\over v} \; I(0,m) + \nonumber\\
& & -{9\over8}{\hbar^2\over v^3} \; I(0,\mu)^2 - {85\over8}{\hbar^2\over v^3} \; I(0,m)^2 + {99\over4}{\hbar^2\over v^3} \; I(0,\mu)I(0,m) + \nonumber\\
& & -9{\hbar^2m^2\over v^3} \; I(0,\mu,0,\mu)I(0,\mu) + 4{\hbar^2m^2\over v^3} \; I(0,m,0,m)I(0,m) + \nonumber\\
& & 21{\hbar^2m^2\over v^3} \; I(0,\mu,0,\mu)I(0,m) - 7{\hbar^2m^2\over v^3} \; I(0,m,0,m)I(0,\mu) + \nonumber\\
& & -27{\hbar^2m^4\over v^3} \; B_{\mu\mu\mu} - 3{\hbar^2m^4\over v^3} \; B_{\mu mm} + 2{\hbar^2m^4\over v^3} \; B_{m\mu m} + \nonumber\\
& & 3{\hbar^2m^2\over v^3} \; D_{\mu\mu\mu} - 11{\hbar^2m^2\over v^3} \; D_{mm\mu} \nonumber\\[5pt]
\langle\f_2(x)\rangle &=& \langle\eta_2(x)\rangle = 0 \nonumber\\[5pt]
\langle\f_1(x)\f_1(y)\rangle &=& v^2 + 2v\langle\eta_1(x)\rangle + \langle\eta_1(x)\eta_1(y)\rangle \nonumber\\
&=& v^2 - 3\hbar \; I(0,\mu) + \hbar \; I(0,m) + \hbar \; A_\mu(x-y) + \nonumber\\
& & -21{\hbar^2\over v^2} \; I(0,m)^2 + 48{\hbar^2\over v^2} \; I(0,\mu)I(0,m) -18{\hbar^2m^2\over v^2} \; I(0,\mu,0,\mu)I(0,\mu) + \nonumber\\
& & 8{\hbar^2m^2\over v^2} \; I(0,m,0,m)I(0,m) + 42{\hbar^2m^2\over v^2} \; I(0,\mu,0,\mu)I(0,m) + \nonumber\\
& & -14{\hbar^2m^2\over v^2} \; I(0,m,0,m)I(0,\mu) - 54{\hbar^2m^4\over v^2} \; B_{\mu\mu\mu} - 6{\hbar^2m^4\over v^2} \; B_{\mu mm} + \nonumber\\
& & 4{\hbar^2m^4\over v^2} \; B_{m\mu m} + 6{\hbar^2m^2\over v^2} \; D_{\mu\mu\mu} - 22{\hbar^2m^2\over v^2} \; D_{mm\mu} + \nonumber\\
& & 6{\hbar^2m^2\over v^2} \; I(0,\mu) C_{\mu\mu}(x-y) - 14{\hbar^2m^2\over v^2} \; I(0,m) C_{\mu\mu}(x-y) + \nonumber\\
& & 18{\hbar^2m^4\over v^2} \; B_{\mu\mu\mu}(x-y) + 2{\hbar^2m^4\over v^2} \; B_{\mu mm}(x-y) \nonumber\\
\langle\f_2(x)\f_2(y)\rangle &=& \langle\eta_2(x)\eta_2(y)\rangle \nonumber\\
&=& \hbar \; A_m(x-y) + \nonumber\\
& & -14{\hbar^2m^2\over v^2} \; I(0,\mu) C_{mm}(x-y) + 8{\hbar^2m^2\over v^2} \; I(0,m) C_{mm}(x-y) + \nonumber\\
& & 4{\hbar^2m^4\over v^2} \; B_{mm\mu}(x-y) \nonumber\\[5pt]
\langle\f_1(x)\f_2(y)\rangle &=& v\langle\eta_2(y)\rangle + \langle\eta_1(x)\eta_2(y)\rangle = 0 \label{Cartarctan}
\eqa}

\subsection{Polar Results}

According to the conjecture we can just transform the continuum action (\ref{Cartactionarctan}) to an action in terms of the polar field variables to obtain the Feynman rules for the polar calculation. So the action becomes:
\bq
S = \int d^dx \left( \h\left(\nabla r\right)^2 + \h{r^2\over v^2}\left(\nabla w\right)^2 + {m^2\over4v^2}(r^2-v^2)^2 + \h m^2v^2{w^2\over r^2} \right)
\eq
This can be expanded around $r=v$ again. Defining
\bq
r(x) \equiv v + \eta(x) \;,
\eq
we find the following Feynman rules for the $\eta$- and $w$-field (up to order $\hbar^{5/2}$). The vertices from the Jacobian are exactly the same as in the shifted toy model.
{\allowdisplaybreaks\bqa
\begin{picture}(100, 20)(0, 17)
\Line(20, 20)(80, 20)
\end{picture}
&\leftrightarrow& \frac{\hbar}{k^2+\mu^2} \nonumber\\
\begin{picture}(100, 20)(0, 17)
\DashLine(20, 20)(80, 20){5}
\end{picture}
&\leftrightarrow& \frac{\hbar}{k^2+m^2} \nonumber\\
\begin{picture}(100, 40)(0, 17)
\Line(20, 20)(50, 20)
\Line(50, 20)(80, 0)
\Line(50, 20)(80, 40)
\end{picture}
&\leftrightarrow& -\frac{1}{\hbar}\frac{6m^2}{v} \nonumber\\
\begin{picture}(100, 40)(0, 17)
\Line(20, 20)(50, 20)
\DashLine(50, 20)(80, 0){5}
\DashLine(50, 20)(80, 40){5}
\Text(80,30)[c]{$k_1$}
\Text(80,10)[c]{$k_2$}
\end{picture}
&\leftrightarrow& \frac{2}{\hbar v} k_1\cdot k_2 + \frac{1}{\hbar}\frac{2m^2}{v} \nonumber\\
\begin{picture}(100, 40)(0, 17)
\Line(20, 0)(80, 40)
\Line(20, 40)(80, 0)
\end{picture}
&\leftrightarrow& -\frac{1}{\hbar}\frac{6m^2}{v^2} \nonumber\\
\begin{picture}(100, 40)(0, 17)
\Line(20,0)(50,20)
\Line(20,40)(50,20)
\DashLine(50,20)(80,40){5}
\DashLine(50,20)(80,0){5}
\Text(80,30)[c]{$k_1$}
\Text(80,10)[c]{$k_2$}
\end{picture}
&\leftrightarrow& \frac{2}{\hbar v^2} k_1\cdot k_2 - \frac{1}{\hbar}\frac{6m^2}{v^2} \nonumber\\
\begin{picture}(100, 40)(0, 17)
\Line(20, 0)(50, 20)
\Line(20, 20)(50, 20)
\Line(20, 40)(50, 20)
\DashLine(50, 20)(80, 40){5}
\DashLine(50, 20)(80, 0){5}
\end{picture}
&\leftrightarrow& \frac{1}{\hbar}\frac{24m^2}{v^3} \nonumber\\
\begin{picture}(100, 40)(0, 17)
\Line(20, 20)(50, 20)
\Vertex(50, 20){3}
\end{picture}
&\leftrightarrow& {1\over v} \; I \nonumber\\
\begin{picture}(100, 40)(0, 17)
\Line(20, 20)(80, 20)
\Vertex(50, 20){3}
\end{picture}
&\leftrightarrow& -{1\over v^2} \; I \nonumber\\
\begin{picture}(100, 40)(0, 17)
\Line(20, 20)(50, 20)
\Line(50, 20)(80, 40)
\Line(50, 20)(80, 0)
\Vertex(50, 20){3}
\end{picture}
&\leftrightarrow& {2\over v^3} \; I
\eqa}\\
Here the solid lines denote the $\eta$-field, the dashed lines denote the $w$-field and all momenta are counted into the vertex. Also we have defined $\mu=\sqrt{2}m$, as in the Cartesian calculation.

Now we can again compute the $\eta$- and $w$-Green's-functions up to order $\hbar^2$.

\subsubsection{The $\eta$-Tadpole}

For the $\eta$-tadpole we find the following result, up to order $\hbar^2$. For a more detailed calculation we refer to \cite{vanKessel}.
\vspace{-23pt}
{\allowdisplaybreaks\bqa
\langle\tilde{\eta}\rangle &=&
\begin{picture}(100,50)(0,23)
\Line(0,25)(50,25)
\CArc(75,25)(25,0,360)
\end{picture} \quad + \quad
\begin{picture}(100,50)(0,23)
\Line(0,25)(50,25)
\DashCArc(75,25)(25,0,360){3}
\end{picture} \quad + \quad
\begin{picture}(100,50)(0,23)
\Line(0,25)(75,25)
\Vertex(75,25){4}
\end{picture} \quad + \nonumber\\[10pt]
& & \begin{picture}(100,50)(0,23)
\Line(0,25)(35,25)
\CArc(60,25)(25,0,360)
\GCirc(85,25){15}{0.5}
\end{picture} \quad + \quad
\begin{picture}(100,50)(0,23)
\Line(0,25)(35,25)
\DashCArc(60,25)(25,0,360){3}
\GCirc(85,25){15}{0.5}
\end{picture} \quad + \quad
\begin{picture}(100,50)(0,23)
\Line(0,25)(60,25)
\CArc(30,40)(15,0,360)
\GCirc(80,25){20}{0.5}
\end{picture} \quad + \nonumber\\[10pt]
& & \begin{picture}(100,50)(0,23)
\Line(0,25)(60,25)
\DashCArc(30,40)(15,0,360){3}
\GCirc(80,25){20}{0.5}
\end{picture} \quad + \quad
\begin{picture}(100,50)(0,23)
\Line(0,25)(60,25)
\Vertex(30,25){4}
\GCirc(80,25){20}{0.5}
\end{picture} \quad + \quad
\begin{picture}(100,50)(0,23)
\Line(0,25)(50,25)
\Line(50,25)(85,42)
\Line(50,25)(85,8)
\GCirc(85,42){15}{0.5}
\GCirc(85,8){15}{0.5}
\end{picture} \quad + \nonumber\\[10pt]
& & \begin{picture}(100,50)(0,23)
\Line(0,25)(100,25)
\CArc(75,25)(25,0,360)
\end{picture} \quad + \quad
\begin{picture}(100,50)(0,23)
\Line(0,25)(100,25)
\DashCArc(75,25)(25,0,360){3}
\end{picture} \quad + \quad
\begin{picture}(100,50)(0,23)
\Line(0,25)(75,25)
\CArc(75,37.5)(12.5,0,360)
\DashCArc(75,12.5)(12.5,0,360){3}
\end{picture} \quad + \nonumber\\[10pt]
& & \begin{picture}(100,50)(0,23)
\Line(0,25)(50,25)
\CArc(75,25)(25,0,360)
\Vertex(50,25){4}
\end{picture} \nonumber\\[30pt]
&=& -{3\over2}{\hbar\over v} \; I(0,\mu) + {\hbar\over v} \; I(0,m) \nonumber\\
& & -{9\over8}{\hbar^2\over v^3} \; I(0,\mu)^2 - {45\over4}{\hbar^2\over v^3} \; I(0,m)^2 + 26{\hbar^2\over v^3} \; I(0,\mu) I(0,m) \nonumber\\
& & -9{\hbar^2m^2\over v^3} \; I(0,\mu,0,\mu) I(0,\mu) + 8{\hbar^2m^2\over v^3} \; I(0,m,0,m) I(0,m) \nonumber\\
& & +21{\hbar^2m^2\over v^3} \; I(0,\mu,0,\mu) I(0,m) - 14{\hbar^2m^2\over v^3} \; I(0,m,0,m) I(0,\mu) \nonumber\\
& & -27{\hbar^2m^4\over v^3} \; B_{\mu\mu\mu} - 3{\hbar^2m^4\over v^3} \; B_{\mu mm} + 4{\hbar^2m^4\over v^3} \; B_{m\mu m} \nonumber\\
& & +3{\hbar^2m^2\over v^3} \; D_{\mu\mu\mu} - 12{\hbar^2m^2\over v^3} \; D_{mm\mu}
\eqa}
Notice that, as in the shifted toy model, also here the badly divergent $I$-integrals cancel.

\subsubsection{The $\eta$-Propagator}

The momentum-space $\eta$-propagator is:
\vspace{-18pt}
{\allowdisplaybreaks\bqa
\langle \tilde{\eta}(p)\tilde{\eta}(-p) \rangle_{\mathrm{c}} &=&
\begin{picture}(70,40)(0,18)
\Line(0,20)(70,20)
\end{picture} \quad + \quad
\begin{picture}(70,40)(0,18)
\Line(0,20)(20,20)
\CArc(35,20)(15,-180,180)
\Line(50,20)(70,20)
\end{picture} \quad + \quad
\begin{picture}(70,40)(0,18)
\Line(0,20)(20,20)
\DashCArc(35,20)(15,-180,180){3}
\Line(50,20)(70,20)
\end{picture} \quad + \nonumber\\
& & \begin{picture}(70,40)(0,18)
\Line(0,5)(70,5)
\CArc(35,20)(15,-90,270)
\end{picture} \quad + \quad
\begin{picture}(70,40)(0,18)
\Line(0,5)(70,5)
\DashCArc(35,20)(15,-90,270){3}
\end{picture} \quad + \quad
\begin{picture}(70,40)(0,18)
\Line(0,5)(70,5)
\Line(35,5)(35,25)
\GCirc(35,25){10}{0.5}
\end{picture} \quad + \nonumber\\
& & \begin{picture}(70,40)(0,18)
\Line(0,20)(70,20)
\Vertex(35,20){4}
\end{picture} \nonumber\\[10pt]
&=& \frac{\hbar}{p^2+\mu^2} + 6{\hbar^2m^2\over v^2}\frac{1}{(p^2+\mu^2)^2} \; I(0,\mu) - 14{\hbar^2m^2\over v^2}\frac{1}{(p^2+\mu^2)^2} \; I(0,m) + \nonumber\\
& & -{\hbar^2\over v^2}\frac{1}{p^2+\mu^2} \; I(0,m) + 18{\hbar^2m^4\over v^2}\frac{1}{(p^2+\mu^2)^2} \; I(0,\mu,p,\mu) + \nonumber\\
& & 2{\hbar^2m^4\over v^2}\frac{1}{(p^2+\mu^2)^2} \; I(0,m,p,m) + 2{\hbar^2m^2\over v^2}\frac{1}{p^2+\mu^2} \; I(0,m,p,m) + \nonumber\\
& & \h{\hbar^2\over v^2} \; I(0,m,p,m)
\eqa}

\subsubsection{The $w$-Propagator}

The $w$-propagator up to order $\hbar^2$ is:
\vspace{-18pt}
{\allowdisplaybreaks\bqa
\langle\tilde{w}(p)\tilde{w}(-p)\rangle_c &=&
\begin{picture}(70,40)(0,18)
\DashLine(0,20)(70,20){3}
\end{picture} \quad + \quad
\begin{picture}(70,40)(0,18)
\DashLine(0,20)(20,20){3}
\DashCArc(35,20)(15,-180,0){3}
\CArc(35,20)(15,0,180)
\DashLine(50,20)(70,20){3}
\end{picture} \quad + \quad
\begin{picture}(70,40)(0,18)
\DashLine(0,5)(70,5){3}
\Line(35,5)(35,25)
\GCirc(35,25){10}{0.5}
\end{picture} \quad + \nonumber\\
& & \begin{picture}(70,40)(0,18)
\DashLine(0,5)(70,5){3}
\CArc(35,20)(15,-90,270)
\end{picture} \nonumber\\[20pt]
&=& \frac{\hbar}{p^2+m^2} + 8{\hbar^2m^2\over v^2}\frac{1}{(p^2+m^2)^2} \; I(0,m) - 3{\hbar^2\over v^2}\frac{1}{p^2+m^2} \; I(0,m) + \nonumber\\
& & -14{\hbar^2m^2\over v^2}\frac{1}{(p^2+m^2)^2} \; I(0,\mu) + 5{\hbar^2\over v^2}\frac{1}{p^2+m^2} \; I(0,\mu) + {\hbar^2\over v^2} \; I(0,\mu,p,m) \nonumber\\
& & -4{\hbar^2m^2\over v^2}\frac{1}{p^2+m^2} \; I(0,\mu,p,m) + 4{\hbar^2m^4\over v^2}\frac{1}{(p^2+m^2)^2} \; I(0,\mu,p,m)
\eqa}

The higher-point Green's functions are easily calculated up to order $\hbar^2$, they contain at most tree diagrams.

To compute the Cartesian $\f_1$- and $\f_2$-Green's-functions we can again use the expansions (\ref{Cartaverages}), since these expansions are model independent. First we have to find the configuration-space $\eta$- and $w$-Green's functions from the momentum-space Green's functions above. Then these results can be substituted in (\ref{Cartaverages}). Doing this one finds again the results (\ref{Cartarctan}). So also in case of the arctangent toy model the conjecture is verified for several Green's functions up to order $\hbar^2$. For a more detailed calculation we refer to \cite{vanKessel} again.

\section{Proof of the Conjecture}\label{proofconj}

In the last two sections evidence for the truth of the conjecture has accumulated. In this section we shall prove this conjecture for a general model in $d$ space-time dimensions. 

Our proof will be based on the fact that the transformation to polar field variables actually has to be performed in the path integral on the lattice, i.e.\ in (\ref{genericpathintdiscr}). After transforming to polar fields one gets a path integral in terms of polar fields formulated on a lattice. This path integral gives a (complicated) set of Feynman rules, and diagrams actually have to be calculated with space-time still discrete. Only in the end result for the Green's function one should then take the continuum limit, i.e.\ $\Delta\rightarrow0$.

Now a $d$-dimensional continuum calculation is correct if one can see that all the steps one performs there to calculate a diagram correspond to a similar step in a discrete calculation. In a $d$-dimensional continuum calculation one performs the following three steps when calculating any diagram:
\begin{enumerate}
\item One writes momentum-dependent factors from the vertices in terms of the denominators of propagators, such that one can let them cancel. For example:
\bq
k\cdot l = \h((k+l)^2+m^2) - \h(k^2+m^2) - \h(l^2+m^2) + \h m^2 \;.
\eq
\item One shifts momenta, for example:
\bq
\int d^dk \frac{k^2+m^2}{(k+l)^2+m^2} = \int d^dk \frac{(k-l)^2+m^2}{k^2+m^2} \;.
\eq
\item When there is momentum dependence left in the numerator, which cannot cancel anything in the denominator anymore, one uses
\bq
\int d^dk \; \frac{k_i}{k^2+m^2} = 0 \;.
\eq
For example:
\bqa
& & \int d^dk \int d^dl \; \frac{(k-l)^2+m^2}{(k^2+m^2)(l^2+m^2)} \nonumber\\
& & = \int d^dk \int d^dl \; \frac{(k^2+m^2)+(l^2+m^2)-m^2}{(k^2+m^2)(l^2+m^2)} \nonumber\\
& & = \int d^dk \int d^dl \; \left( \frac{1}{l^2+m^2} + \frac{1}{k^2+m^2} - \frac{m^2}{(k^2+m^2)(l^2+m^2)} \right)
\eqa
\end{enumerate}
If we can somehow see that these steps are also valid in a discrete calculation, then we have proven the conjecture. For then we know that every operation one performs in the $d$-dimensional continuum calculation corresponds to a valid operation in a discrete calculation, even though one writes down these steps in a continuum formalism.

The first step is to transform the discrete action (\ref{discraction}) to polar field variables. The transformation goes as follows:
\bqa
\f_{1 \;\; i_1,\ldots,i_d} &=& r_{i_1,\ldots,i_d} \cos\left(\frac{w_{i_1,\ldots,i_d}}{v}\right) \nonumber\\
\f_{2 \;\; i_1,\ldots,i_d} &=& r_{i_1,\ldots,i_d} \sin\left(\frac{w_{i_1,\ldots,i_d}}{v}\right) \label{discrtransf}
\eqa
To keep things readable we define the shorthand notations:
\bqa
i &\equiv& i_1,\ldots,i_d \nonumber\\
\nabla r_{i_1,\ldots,i_d} &\equiv& \threev{{1\over\Delta}\left(r_{i_1+1,i_2,\ldots,i_d}-r_{i_1,i_2,\ldots,i_d}\right)} {\vdots} {{1\over\Delta}\left(r_{i_1,\ldots,i_{d-1},i_d+1}-r_{i_1,\ldots,i_{d-1},i_d}\right)} \label{shorthand}
\eqa
The action becomes:
\bqa
S &=& \Delta^d \sum_{i_1,\ldots,i_d=0}^{N-1} \Bigg( \h\left(\nabla r_i\right)^2 + {1\over\Delta^2}r_i^2\left( \left(1-\cos\frac{\Delta\nabla_1w_i}{v}\right) + \ldots + \left(1-\cos\frac{\Delta\nabla_dw_i}{v}\right) \right) + \nonumber\\
& & \phantom{\Delta^d \sum_{i_1,\ldots,i_d=0}^{N-1} \Bigg(} {1\over\Delta}r_i\nabla_1r_i \left(1-\cos\frac{\Delta\nabla_1w_i}{v}\right) + \ldots + {1\over\Delta}r_i\nabla_dr_i \left(1-\cos\frac{\Delta\nabla_dw_i}{v}\right) + \nonumber\\
& & \phantom{\Delta^d \sum_{i_1,\ldots,i_d=0}^{N-1} \Bigg(} V\left(r_i\cos{w_i\over v}, r_i\sin{w_i\over v}\right) \Bigg)
\eqa

Now we will substitute the series expansion for the cosines in the first two lines. Also we will assume that the potential is such that the minimum of the complete action is at $r=v$, where $v$ is some nonzero constant (the same $v$ that divides $w$ in the cosine). This assumption is necessary to avoid difficulties with the singularity at $r=0$ in the transformation (\ref{discrtransf}). Because the minimum of the action is at $r=v$ we also express the action in terms of $\eta$, which denotes the deviation from $v$:
\bq
r_i = v+\eta_i \;.
\eq
The final form of the discrete action is:
\bqa
S &=& \Delta^d \sum_{i_1,\ldots,i_d=0}^{N-1} \Bigg( \h\left(\nabla\eta_i\right)^2 + \h\left(\nabla w_i\right)^2 \nonumber\\
& & \phantom{\Delta^d \sum_{i_1,\ldots,i_d=0}^{N-1} \Bigg(} -\sum_{n=2}^\infty \frac{(-1)^n}{(2n)!} \frac{\Delta^{2n-2}}{v^{2n-2}} \left( \left(\nabla_1w_i\right)^{2n} + \ldots + \left(\nabla_dw_i\right)^{2n} \right) \nonumber\\
& & \phantom{\Delta^d \sum_{i_1,\ldots,i_d=0}^{N-1} \Bigg(} -2\sum_{n=1}^\infty \frac{(-1)^n}{(2n)!} \frac{\Delta^{2n-2}}{v^{2n-2}} \left( \eta_i\left(\nabla_1w_i\right)^{2n} + \ldots + \eta_i\left(\nabla_dw_i\right)^{2n} \right) \nonumber\\
& & \phantom{\Delta^d \sum_{i_1,\ldots,i_d=0}^{N-1} \Bigg(} -\sum_{n=1}^\infty \frac{(-1)^n}{(2n)!} \frac{\Delta^{2n-2}}{v^{2n-2}} \left( \left(\nabla_1\eta_i\right)\left(\nabla_1w_i\right)^{2n} + \ldots + \left(\nabla_d\eta_i\right)\left(\nabla_dw_i\right)^{2n} \right) \nonumber\\
& & \phantom{\Delta^d \sum_{i_1,\ldots,i_d=0}^{N-1} \Bigg(} -\sum_{n=1}^\infty \frac{(-1)^n}{(2n)!} \frac{\Delta^{2n-2}}{v^{2n-2}} \left( \eta_i^2\left(\nabla_1w_i\right)^{2n} + \ldots + \eta_i^2\left(\nabla_dw_i\right)^{2n} \right) \nonumber\\
& & \phantom{\Delta^d \sum_{i_1,\ldots,i_d=0}^{N-1} \Bigg(} -\sum_{n=1}^\infty \frac{(-1)^n}{(2n)!} \frac{\Delta^{2n-2}}{v^{2n-2}} \left( \eta_i\left(\nabla_1\eta_i\right)\left(\nabla_1w_i\right)^{2n} + \ldots + \eta_i\left(\nabla_d\eta_i\right)\left(\nabla_dw_i\right)^{2n} \right) \nonumber\\
& & \phantom{\Delta^d \sum_{i_1,\ldots,i_d=0}^{N-1} \Bigg(} +V\left((v+\eta_i)\cos{w_i\over v}, (v+\eta_i)\sin{w_i\over v}\right) \Bigg) \label{discractioncompl}
\eqa

Note that this expression is still exact, as long as we keep all the terms in the sums coming from the expansion of the cosines. Our complete discrete path integral $P$ (defined in (\ref{genericpathint})), formulated in terms of the polar fields, now looks like:
\bq
P = \int_{-v}^\infty \left(\prod_{i=0}^{N-1} (v+\eta_i)d\eta_i\right) \int_{-v\pi}^{v\pi} \left(\prod_{i=0}^{N-1} dw_i\right) \; O \; \exp\left(-{1\over\hbar}S\right) \;,
\eq
where the action $S$ is given by (\ref{discractioncompl}) and $O$ is the product of the $\f_1$- and $\f_2$-fields:
\bq
O \equiv \left(v+\eta_{j^{(1)}}\right)\cos{w_{j^{(1)}}\over v} \cdots \left(v+\eta_{j^{(m)}}\right)\cos{w_{j^{(m)}}\over v} \cdot \left(v+\eta_{k^{(1)}}\right)\sin{w_{k^{(1)}}\over v} \cdots \left(v+\eta_{k^{(n)}}\right)\sin{w_{k^{(n)}}\over v} \;.
\eq 

The product
\bq
\prod_{i=0}^{N-1} (v+\eta_i) = \prod_{i=0}^{N-1} r_i
\eq
is the Jacobian from the transformation to polar fields. This Jacobian factor can be recast in the following form:
\bq
\left(\prod_{i=0}^{N-1} (v+\eta_i)\right) = \exp\left(\sum_{i=0}^{N-1} \ln(v+\eta_i)\right)
\eq

Also the domain of integration for the $w$-fields can be extended from $[-v\pi,v\pi]$ to $\langle-\infty,\infty\rangle$, because the whole integrand is periodic in the $w$-fields. Finally we can also extend the lower integration boundary for the $\eta$-fields from $-v$ to $-\infty$, this shift will only have non-perturbative effects.

So we have brought our path integral to the form
\bq
P = \int_{-\infty}^\infty \left(\prod_{i=0}^{N-1} d\eta_i\right) \int_{-\infty}^{\infty} \left(\prod_{i=0}^{N-1} dw_i\right) \; O \; \exp\left(-{1\over\hbar}S + \sum_{i=0}^{N-1} \ln(v+\eta_i)\right) \;.
\eq
This is a normal path integral on a lattice, in the sense that it has the same form of a path integral in terms of Cartesian fields. Such a path integral we can calculate in the ordinary way, with perturbation theory. The action of this path integral does have an infinite number of vertices however, because the discrete action (\ref{discractioncompl}) has an infinite number of interaction terms and also because the expansion of the logarithm coming from the Jacobian has an infinite number of terms. But it is still an \emph{exact} expression, because we keep \emph{all} terms.

Now we have to write down the (discrete) Feynman rules for the action we have found. To write down the momentum-space Feynman rules we must first transform the configuration-space fields $\eta$ and $w$ to momentum-space fields $\tilde{\eta}$ and $\tilde{w}$. In the continuum such a transformation is given by:
\bq
\tilde{\eta}(k) = \int d^dx \; \eta(x) \; e^{ik\cdot x} \;,
\eq
and similar for the $\tilde{w}$-field. The discrete analogue of this formula is:
\bq \label{discrFourier}
\tilde{\eta}_{k_1,\ldots,k_d} = \Delta^d \sum_{i_1,\ldots,i_d=0}^{N-1} \eta_{i_1,\ldots,i_d} \; \exp\left({2\pi i\over L}\left(k_1(\Delta i_1-L/2)+\ldots+k_d(\Delta i_d-L/2)\right)\right) \;,
\eq
where we have used that the continuous momentum is related to the discrete momentum as
\bq \label{discrcontmom}
k_{\mathrm{cont}} = \frac{2\pi k_{\mathrm{discr}}}{L} \;.
\eq
The inverse transformation of (\ref{discrFourier}) is
\bq \label{invdiscrFourier}
\eta_{i_1,\ldots,i_d} = {1\over L^d} \sum_{k_1,\ldots,k_d=-N/2}^{N/2-1} \tilde{\eta}_{k_1,\ldots,k_d} \; \exp\left(-{2\pi i\over L}\left(k_1(\Delta i_1-L/2)+\ldots+k_d(\Delta i_d-L/2)\right)\right) \;,
\eq
and similar for the $w$-field. To see that this is indeed the inverse transformation of (\ref{discrFourier}) one can use the identity
\bq
{1\over N^d} \sum_{i_1,\ldots,i_d=0}^{N-1} \exp\left({2\pi i\over N}(k_1i_1+\ldots+k_di_d)\right) = \delta_{k_1,0\;\mathrm{mod}\;N} \cdots \delta_{k_d,0\;\mathrm{mod}\;N} \;.
\eq

From the relation (\ref{discrcontmom}) we can see that when we take $L\rightarrow\infty$ the momenta become a continuous set. Their domain is still finite however. Because the discrete momenta are between $-N/2$ and $N/2-1$, the continuous momenta are in the domain
\bq
k_{\mathrm{cont}} \in \left\langle -{\pi\over\Delta}, {\pi\over\Delta} \right\rangle \;.
\eq
The finiteness of this domain reflects the discreteness of space-time. From now on we shall understand that we have taken the limit $L\rightarrow\infty$, such that all sums over momenta become integrals. But of course $\Delta$ is still finite.

By using (\ref{invdiscrFourier}), in the limit $L\rightarrow\infty$, we can now express the discrete action (\ref{discractioncompl}) in terms of the momentum-space fields $\tilde{\eta}$ and $\tilde{w}$. From this action one can then read of the discrete, momentum-space Feynman rules. Notice that we did not specify the potential $V$, so we will not include the Feynman rules coming from this part of the action. This potential $V$ will also determine the masses for the $\eta$- and $w$-field. We shall keep these masses general, the upcoming proof for the conjecture will not depend on the explicit form of the potential $V$ and the masses. In the Feynman rules below we will neither include the Feynman rules from the Jacobian, the proof of the conjecture will also not depend on the exact form of these vertices.

The discrete, momentum-space Feynman rules are then:
{\allowdisplaybreaks\bqa
\begin{picture}(100, 20)(0, 17)
\Line(20, 20)(80, 20)
\end{picture}
&\leftrightarrow& \frac{\hbar}{{2d\over\Delta^2}-{2\over\Delta^2}\cos\Delta k_1-\ldots-{2\over\Delta^2}\cos\Delta k_d+m_\eta^2} \nonumber\\
\begin{picture}(100, 20)(0, 17)
\DashLine(20, 20)(80, 20){5}
\end{picture}
&\leftrightarrow& \frac{\hbar}{{2d\over\Delta^2}-{2\over\Delta^2}\cos\Delta k_1-\ldots-{2\over\Delta^2}\cos\Delta k_d+m_w^2} \nonumber\\
\begin{picture}(100, 40)(0, 17)
\Line(20, 20)(50, 20)
\DashLine(50, 20)(80, 0){5}
\DashLine(50, 20)(80, 40){5}
\Text(20, 26)[]{$p$}
\Text(84, 32)[]{$k^{(1)}$}
\Text(84, 10)[]{$k^{(2)}$}
\end{picture}
&\leftrightarrow& -\frac{1}{\hbar v}{1\over\Delta^2} \bigg[ \left(e^{-i\Delta p_1}+1\right)\left(e^{-i\Delta k^{(1)}_1}-1\right)\left(e^{-i\Delta k^{(2)}_1}-1\right) \nonumber\\
& & \phantom{-\frac{1}{\hbar v}{1\over\Delta^2} \bigg[} + \ldots + \nonumber\\
& & \phantom{-\frac{1}{\hbar v}{1\over\Delta^2} \bigg[} \left(e^{-i\Delta p_d}+1\right)\left(e^{-i\Delta k^{(1)}_d}-1\right)\left(e^{-i\Delta k^{(2)}_d}-1\right) \bigg] \nonumber\\
&\vdots& \nonumber\\
\begin{picture}(100, 40)(0, 17)
\Line(20, 20)(50, 20)
\DashLine(50, 20)(80, 0){5}
\DashLine(50, 20)(80, 40){5}
\DashCArc(50,20)(30,-20,20){2}
\Text(20, 26)[]{$p$}
\Text(84, 48)[]{$k^{(1)}$}
\Text(84, -8)[]{$k^{(2n)}$}
\end{picture}
&\leftrightarrow& \frac{(-1)^n}{\hbar v^{2n-1}}{1\over\Delta^2} \bigg[ \left(e^{-i\Delta p_1}+1\right)\left(e^{-i\Delta k^{(1)}_1}-1\right) \cdots \left(e^{-i\Delta k^{(2n)}_1}-1\right) \nonumber\\
& & \phantom{\frac{(-1)^n}{\hbar v^{2n-1}}{1\over\Delta^2} \bigg[} + \ldots + \nonumber\\
& & \phantom{\frac{(-1)^n}{\hbar v^{2n-1}}{1\over\Delta^2} \bigg[} \left(e^{-i\Delta p_d}+1\right)\left(e^{-i\Delta k^{(1)}_d}-1\right) \cdots \left(e^{-i\Delta k^{(2n)}_d}-1\right) \bigg] \nonumber\\
&\vdots& \nonumber\\
\begin{picture}(100, 40)(0, 17)
\Line(20,0)(50,20)
\Line(20,40)(50,20)
\DashLine(50, 20)(80, 0){5}
\DashLine(50, 20)(80, 40){5}
\Text(20, 32)[]{$p$}
\Text(20, 10)[]{$q$}
\Text(84, 32)[]{$k^{(1)}$}
\Text(84, 10)[]{$k^{(2)}$}
\end{picture}
&\leftrightarrow& -\frac{1}{\hbar v^2}{1\over\Delta^2} \bigg[ \left(e^{-i\Delta p_1}+e^{-i\Delta q_1}\right)\left(e^{-i\Delta k^{(1)}_1}-1\right)\left(e^{-i\Delta k^{(2)}_1}-1\right) \nonumber\\
& & \phantom{-\frac{1}{\hbar v^2}{1\over\Delta^2} \bigg[} + \ldots + \nonumber\\
& & \phantom{-\frac{1}{\hbar v^2}{1\over\Delta^2} \bigg[} \left(e^{-i\Delta p_d}+e^{-i\Delta q_d}\right)\left(e^{-i\Delta k^{(1)}_d}-1\right)\left(e^{-i\Delta k^{(2)}_d}-1\right) \bigg] \nonumber\\
&\vdots& \nonumber\\
\begin{picture}(100, 40)(0, 17)
\Line(20,0)(50,20)
\Line(20,40)(50,20)
\DashLine(50, 20)(80, 0){5}
\DashLine(50, 20)(80, 40){5}
\DashCArc(50,20)(30,-20,20){2}
\Text(20, 32)[]{$p$}
\Text(20, 10)[]{$q$}
\Text(84, 48)[]{$k^{(1)}$}
\Text(84, -8)[]{$k^{(2n)}$}
\end{picture}
&\leftrightarrow& \frac{(-1)^n}{\hbar v^{2n}}{1\over\Delta^2} \bigg[ \left(e^{-i\Delta p_1}+e^{-i\Delta q_1}\right)\left(e^{-i\Delta k^{(1)}_1}-1\right) \cdots \left(e^{-i\Delta k^{(2n)}_1}-1\right) \nonumber\\
& & \phantom{\frac{(-1)^n}{\hbar v^{2n}}{1\over\Delta^2} \bigg[} + \ldots + \nonumber\\
& & \phantom{\frac{(-1)^n}{\hbar v^{2n}}{1\over\Delta^2} \bigg[} \left(e^{-i\Delta p_d}+e^{-i\Delta q_d}\right)\left(e^{-i\Delta k^{(1)}_d}-1\right) \cdots \left(e^{-i\Delta k^{(2n)}_d}-1\right) \bigg] \nonumber\\
&\vdots& \nonumber\\
\begin{picture}(100, 40)(0, 17)
\DashLine(20,0)(80,40){5}
\DashLine(20,40)(80,0){5}
\Text(20, 32)[]{$k^{(1)}$}
\Text(20, 10)[]{$k^{(2)}$}
\Text(84, 32)[]{$k^{(3)}$}
\Text(84, 10)[]{$k^{(4)}$}
\end{picture}
&\leftrightarrow& \frac{1}{\hbar v^2}{1\over\Delta^2} \bigg[ \left(e^{-i\Delta k^{(1)}_1}-1\right) \cdots \left(e^{-i\Delta k^{(4)}_1}-1\right) \nonumber\\
& & \phantom{\frac{1}{\hbar v^2}{1\over\Delta^2} \bigg[} + \ldots + \nonumber\\
& & \phantom{\frac{1}{\hbar v^2}{1\over\Delta^2} \bigg[} \left(e^{-i\Delta k^{(1)}_d}-1\right) \cdots \left(e^{-i\Delta k^{(4)}_d}-1\right) \bigg] \nonumber\\
&\vdots& \nonumber\\
\begin{picture}(100, 40)(0, 17)
\DashLine(50,20)(80,40){5}
\DashLine(50,20)(80,0){5}
\DashCArc(50,20)(30,50,310){2}
\Text(84, 48)[]{$k^{(1)}$}
\Text(84, -8)[]{$k^{(2n)}$}
\end{picture}
&\leftrightarrow& \frac{(-1)^n}{\hbar v^{2n-2}}{1\over\Delta^2} \bigg[ \left(e^{-i\Delta k^{(1)}_1}-1\right) \cdots \left(e^{-i\Delta k^{(2n)}_1}-1\right) \nonumber\\
& & \phantom{\frac{(-1)^n}{\hbar v^{2n-2}}{1\over\Delta^2} \bigg[} + \ldots + \nonumber\\
& & \phantom{\frac{(-1)^n}{\hbar v^{2n-2}}{1\over\Delta^2} \bigg[} \left(e^{-i\Delta k^{(1)}_d}-1\right) \cdots \left(e^{-i\Delta k^{(2n)}_d}-1\right) \bigg] \nonumber\\
&\vdots& \label{discrFeynman}
\eqa}
Here all the (continuum) momenta are counted incoming. Together with these Feynman rules for the propagator and the vertices we have the rule that every internal momentum should be integrated over from $-\pi/\Delta$ to $\pi/\Delta$.

Now all these vertices can be written in a more convenient form. By combining all the complex exponentials (i.e.\ writing out all products) and using momentum conservation at the vertex one will notice that for each exponential also its complex conjugate occurs. They can be combined into a cosine, the same cosine that occurs in the discrete propagator. In this way we can write the vertex expressions above in terms of the denominator of the propagator. As a shorthand notation we denote the denominators by $\Pi_k$ and $\bar{\Pi}_k$:
\bqa
\Pi_k &\equiv& {2d\over\Delta^2}-{2\over\Delta^2}\cos{2\pi k_1\over N}-\ldots-{2\over\Delta^2}\cos{2\pi k_d\over N}+m_w^2 \nonumber\\
\bar{\Pi}_k &\equiv& {2d\over\Delta^2}-{2\over\Delta^2}\cos{2\pi k_1\over N}-\ldots-{2\over\Delta^2}\cos{2\pi k_d\over N}+m_\eta^2
\eqa
To write the vertices into this more convenient form we also have to define an operator $P$. We denote the set consisting of the $j^\mathrm{th}$ components of the momenta $k^{(1)},\ldots,k^{(2n)}$ by $\{k_j\}$:
\bq
\{k_j\} \equiv \{k_j^{(1)},k_j^{(2)},\ldots,k_j^{(2n)}\} \;.
\eq
Then the operator $P_i$ working on $\{k_j\}$ returns the sum of $i$ momenta chosen from the set $\{k_j\}$. There are ${2n\choose i}$ ways to choose $i$ momenta from a set of $2n$ momenta, so there are also ${2n\choose i}$ different operators $P_i$. For example:
\bq
P_2 \{k_j\} = k_j^{(1)} + k_j^{(2)} \;.
\eq
With these notations we can write the vertex expressions as follows.
{\allowdisplaybreaks\bqa
\begin{picture}(100, 40)(0, 17)
\Line(20, 20)(50, 20)
\DashLine(50, 20)(80, 0){5}
\DashLine(50, 20)(80, 40){5}
\Text(20, 26)[]{$p$}
\Text(84, 32)[]{$k^{(1)}$}
\Text(84, 10)[]{$k^{(2)}$}
\end{picture}
&\leftrightarrow& -\frac{1}{\hbar v}\left( \Pi_{k^{(1)}} + \Pi_{k^{(2)}} - \bar{\Pi}_p - 2m_w^2 + m_\eta^2 \right) \nonumber\\
&\vdots& \nonumber\\
\begin{picture}(100, 40)(0, 17)
\Line(20, 20)(50, 20)
\DashLine(50, 20)(80, 0){5}
\DashLine(50, 20)(80, 40){5}
\DashCArc(50,20)(30,-20,20){2}
\Text(20, 26)[]{$p$}
\Text(84, 48)[]{$k^{(1)}$}
\Text(84, -8)[]{$k^{(2n)}$}
\end{picture}
&\leftrightarrow& \frac{(-1)^n}{\hbar v^{2n-1}} \Bigg( -\bar{\Pi}_p + \sum_{P_1} \Pi_{P_1\{k\}} + \sum_{P_1} \Pi_{p+P_1\{k\}} + \nonumber\\
& & \phantom{\frac{(-1)^n}{\hbar v^{2n-1}} \Bigg(} -\sum_{P_2} \bar{\Pi}_{P_2\{k\}} - \sum_{P_2} \bar{\Pi}_{p+P_2\{k\}} + \ldots + \nonumber\\
& & \phantom{\frac{(-1)^n}{\hbar v^{2n-1}} \Bigg(} (-1)^n\sum_{P_{n-1}} \Pi_{P_{n-1}\{k\}} + (-1)^n\sum_{P_{n-1}} \Pi_{p+P_{n-1}\{k\}} + \nonumber\\
& & \phantom{\frac{(-1)^n}{\hbar v^{2n-1}} \Bigg(} \h(-1)^{n+1}\sum_{P_n} \bar{\Pi}_{P_n\{k\}} + \h(-1)^{n+1}\sum_{P_n} \bar{\Pi}_{p+P_n\{k\}} + \nonumber\\
& & \phantom{\frac{(-1)^n}{\hbar v^{2n-1}} \Bigg(} -2^{2n-1}m_w^2 - \left(1-2^{2n-1}\right)m_\eta^2 \Bigg) \nonumber\\
&\vdots& \nonumber\\
\begin{picture}(100, 40)(0, 17)
\Line(20,0)(50,20)
\Line(20,40)(50,20)
\DashLine(50, 20)(80, 0){5}
\DashLine(50, 20)(80, 40){5}
\Text(20, 32)[]{$p$}
\Text(20, 10)[]{$q$}
\Text(84, 32)[]{$k^{(1)}$}
\Text(84, 10)[]{$k^{(2)}$}
\end{picture}
&\leftrightarrow& -\frac{1}{\hbar v^2}\bigg( -\bar{\Pi}_p - \bar{\Pi}_q + \nonumber\\
& & \phantom{-\frac{1}{\hbar v^2}\bigg(} \h\Pi_{p+k^{(1)}} + \h\Pi_{p+k^{(2)}} + \h\Pi_{q+k^{(1)}} + \h\Pi_{q+k^{(2)}} + \nonumber\\
& & \phantom{-\frac{1}{\hbar v^2}\bigg(} -2m_w^2 + 2m_\eta^2 \bigg) \nonumber\\
&\vdots& \nonumber\\
\begin{picture}(100, 40)(0, 17)
\Line(20,0)(50,20)
\Line(20,40)(50,20)
\DashLine(50, 20)(80, 0){5}
\DashLine(50, 20)(80, 40){5}
\DashCArc(50,20)(30,-20,20){2}
\Text(20, 32)[]{$p$}
\Text(20, 10)[]{$q$}
\Text(84, 48)[]{$k^{(1)}$}
\Text(84, -8)[]{$k^{(2n)}$}
\end{picture}
&\leftrightarrow& \frac{(-1)^n}{\hbar v^{2n}}\Bigg( -\bar{\Pi}_p - \bar{\Pi}_q + \sum_{P_1} \Pi_{p+P_1\{k\}} + \sum_{P_1} \Pi_{q+P_1\{k\}} + \nonumber\\
& & \phantom{\frac{(-1)^n}{\hbar v^{2n}}\Bigg(} -\sum_{P_2} \bar{\Pi}_{p+P_2\{k\}} - \sum_{P_2} \bar{\Pi}_{q+P_2\{k\}} + \ldots + \nonumber\\
& & \phantom{\frac{(-1)^n}{\hbar v^{2n}}\Bigg(} (-1)^n\sum_{P_{n-1}} \Pi_{p+P_{n-1}\{k\}} + (-1)^n\sum_{P_{n-1}} \Pi_{q+P_{n-1}\{k\}} + \nonumber\\
& & \phantom{\frac{(-1)^n}{\hbar v^{2n}}\Bigg(} \h(-1)^{n+1}\sum_{P_n} \bar{\Pi}_{p+P_n\{k\}} + \h(-1)^{n+1}\sum_{P_n} \bar{\Pi}_{q+P_n\{k\}} + \nonumber\\
& & \phantom{\frac{(-1)^n}{\hbar v^{2n}}\Bigg(} - 2^{2n-1}m_w^2 + 2^{2n-1}m_\eta^2 \Bigg) \nonumber\\
&\vdots& \nonumber\\
\begin{picture}(100, 40)(0, 17)
\DashLine(20,0)(80,40){5}
\DashLine(20,40)(80,0){5}
\Text(20, 32)[]{$k^{(1)}$}
\Text(20, 10)[]{$k^{(2)}$}
\Text(84, 32)[]{$k^{(3)}$}
\Text(84, 10)[]{$k^{(4)}$}
\end{picture}
&\leftrightarrow& \frac{1}{\hbar v^2}\bigg( \Pi_{k^{(1)}} + \Pi_{k^{(2)}} + \Pi_{k^{(3)}} + \Pi_{k^{(4)}} + \nonumber\\
& & \phantom{\frac{1}{\hbar v^2}\bigg(} -\h\bar{\Pi}_{k^{(1)}+k^{(2)}} - \h\bar{\Pi}_{k^{(1)}+k^{(3)}} - \h\bar{\Pi}_{k^{(1)}+k^{(4)}} + \nonumber\\
& & \phantom{\frac{1}{\hbar v^2}\bigg(} -\h\bar{\Pi}_{k^{(2)}+k^{(3)}} - \h\bar{\Pi}_{k^{(2)}+k^{(4)}} - \h\bar{\Pi}_{k^{(3)}+k^{(4)}} - 4m_w^2 + 3m_\eta^2 \bigg) \nonumber\\
&\vdots& \nonumber\\
\begin{picture}(100, 40)(0, 17)
\DashLine(50,20)(80,40){5}
\DashLine(50,20)(80,0){5}
\DashCArc(50,20)(30,50,310){2}
\Text(84, 48)[]{$k^{(1)}$}
\Text(84, -8)[]{$k^{(2n)}$}
\end{picture}
&\leftrightarrow& \frac{(-1)^n}{\hbar v^{2n-2}}\Bigg( \sum_{P_1} \Pi_{P_1\{k\}} - \sum_{P_2} \bar{\Pi}_{P_2\{k\}} + \ldots + \nonumber\\
& & \phantom{\frac{(-1)^n}{\hbar v^{2n-2}}\Bigg(} (-1)^n\sum_{P_{n-1}} \Pi_{P_{n-1}\{k\}} + \h(-1)^{n+1}\sum_{P_n} \bar{\Pi}_{P_n\{k\}} \nonumber\\
& & \phantom{\frac{(-1)^n}{\hbar v^{2n-2}}\Bigg(} -2^{2n-2}m_w^2 - \left(1-2^{2n-2}\right)m_\eta^2 \Bigg) \nonumber\\ \label{discrFeynmanrules}
&\vdots&
\eqa}
Notice that the bars are always placed on the terms containing an operator $P_i$ with $i$ even. For the sake of the argument it is convenient to place these bars in this way. Whether or not a bar is placed on the last term, with $P_n$, thus depends on $n$, whether $n$ is even or odd. Above the bars are placed as if $n$ were even, but it should be clear how they should be placed when $n$ is odd.

Now notice that these vertex rules look identical to the rules one would use when doing a $d$-dimensional continuum calculation. For example, in the continuum the 3-vertex would be:
\vspace{-17pt}
\bq
\begin{picture}(100, 40)(0, 17)
\Line(20, 20)(50, 20)
\DashLine(50, 20)(80, 0){5}
\DashLine(50, 20)(80, 40){5}
\Text(20, 26)[]{$p$}
\Text(84, 32)[]{$k^{(1)}$}
\Text(84, 10)[]{$k^{(2)}$}
\end{picture}
\leftrightarrow \frac{2}{\hbar v} k^{(1)}\cdot k^{(2)}
\eq\\
To simplify the dot-product in a $d$-dimensional calculation one would write it as
\bqa
\frac{2}{\hbar v} k^{(1)}\cdot k^{(2)} &=& -\frac{1}{\hbar v}\bigg( \left(\left(k^{(1)}\right)^2+m_w^2\right) + \left(\left(k^{(2)}\right)^2+m_w^2\right) + \nonumber\\
& & \phantom{-\frac{1}{\hbar v}\bigg(} -\left(\left(k^{(1)}+k^{(2)}\right)^2+m_\eta^2\right) - 2m_w^2 + m_\eta^2 \bigg) \;,
\eqa
which corresponds exactly to the discrete vertex expression given in (\ref{discrFeynmanrules}) for the 3-vertex. So when one uses the continuum rules to rewrite dot-products of momenta, as in the 3-vertex example above, one is actually doing a \emph{correct} calculation, although one is doing a continuum calculation.

Another rule that one uses in a continuum calculation is that it is allowed to shift the loop momenta. Also in a discrete calculation this is allowed, because of the periodicity of the discrete propagators and vertex expressions.

What then goes wrong in a continuum calculation? There is one more rule that one uses in a continuum calculation that we have not mentioned up to now. This rule is:
\bq \label{badrule}
{1\over2\pi}\int d^dk \; \frac{k_i}{k^2+m^2} = 0 \;.
\eq
This rule, however, is \emph{not} correct in a discrete calculation. For example, in dimension 1, we have to realize that by $k$ and $k^2+m^2$ we actually mean:
\bqa
k &\leftrightarrow& {i\over\Delta}\left(e^{-i\Delta k}-1\right) \nonumber\\
k^2+m^2 &\leftrightarrow& {2\over\Delta^2}-{2\over\Delta^2}\cos(\Delta k)+m^2
\eqa
So by the integral above we actually mean:
\bq
{1\over2\pi}\int dk \; \frac{k}{k^2+m^2} \leftrightarrow {1\over2\pi}\int_{-\pi/\Delta}^{\pi/\Delta} dk \; \frac{{i\over\Delta}\left(e^{-i\Delta k}-1\right)}{{2\over\Delta^2}-{2\over\Delta^2}\cos(\Delta k)+m^2} = -\h i + \mathcal{O}(\Delta)
\eq
This shows that the rule (\ref{badrule}) is \emph{not} the correct one to use. The only instances that one would use the rule (\ref{badrule}) is when a $\Pi$ (or $\bar{\Pi}$) is left in the numerator, and \emph{cannot} cancel anything in the denominator anymore.

Below we shall show that all such terms, where a $\Pi$ (or $\bar{\Pi}$) remains in the numerator, cancel when one adds \emph{all} diagrams for a certain Green's function. To this end it is convenient to split up the vertices in (\ref{discrFeynmanrules}) as follows:
{\allowdisplaybreaks\bqa
\begin{picture}(100, 40)(0, 17)
\Line(20, 20)(50, 20)
\DashLine(50, 20)(80, 0){5}
\DashLine(50, 20)(80, 40){5}
\DashCArc(50,20)(30,-20,20){2}
\Vertex(40,20){3}
\Text(20, 26)[]{$p$}
\Text(84, 48)[]{$k^{(1)}$}
\Text(84, -8)[]{$k^{(2n)}$}
\end{picture}
&\leftrightarrow& \frac{(-1)^n}{\hbar v^{2n-1}} \left( -\bar{\Pi}_p \right)  \nonumber\\[20pt]
\begin{picture}(100, 40)(0, 17)
\Line(20, 20)(50, 20)
\DashLine(50, 20)(80, 0){5}
\DashLine(50, 20)(80, 40){5}
\DashCArc(50,20)(30,-20,20){2}
\Vertex(58.3,25.5){3}
\Text(20, 26)[]{$p$}
\Text(84, 48)[]{$k^{(1)}$}
\Text(84, -8)[]{$k^{(2n)}$}
\end{picture}
&\leftrightarrow& \frac{(-1)^n}{\hbar v^{2n-1}} \Pi_{k^{(1)}} \nonumber\\[20pt]
&\vdots& \nonumber\\
\begin{picture}(100, 40)(0, 17)
\Line(20, 20)(50, 20)
\DashLine(50, 20)(80, 0){5}
\DashLine(50, 20)(80, 40){5}
\DashCArc(50,20)(30,-20,20){2}
\Vertex(58.3,14.5){3}
\Text(20, 26)[]{$p$}
\Text(84, 48)[]{$k^{(1)}$}
\Text(84, -8)[]{$k^{(2n)}$}
\end{picture}
&\leftrightarrow& \frac{(-1)^n}{\hbar v^{2n-1}} \Pi_{k^{(2n)}} \nonumber\\[20pt]
\begin{picture}(100, 40)(0, 17)
\Line(20, 20)(50, 20)
\DashLine(50, 20)(80, 0){5}
\DashLine(50, 20)(80, 40){5}
\DashCArc(50,20)(30,-20,20){2}
\Vertex(50,20){3}
\Text(20, 26)[]{$p$}
\Text(84, 48)[]{$k^{(1)}$}
\Text(84, -8)[]{$k^{(2n)}$}
\end{picture}
&\leftrightarrow& \frac{(-1)^n}{\hbar v^{2n-1}} \Bigg( \sum_{P_1} \Pi_{p+P_1\{k\}} - \sum_{P_2} \bar{\Pi}_{P_2\{k\}} - \sum_{P_2} \bar{\Pi}_{p+P_2\{k\}} + \ldots + \nonumber\\
& & \phantom{\frac{(-1)^n}{\hbar v^{2n-1}} \Bigg(} (-1)^n\sum_{P_{n-1}} \Pi_{P_{n-1}\{k\}} + (-1)^n\sum_{P_{n-1}} \Pi_{p+P_{n-1}\{k\}} + \nonumber\\
& & \phantom{\frac{(-1)^n}{\hbar v^{2n-1}} \Bigg(} \h(-1)^{n+1}\sum_{P_n} \bar{\Pi}_{P_n\{k\}} + \h(-1)^{n+1}\sum_{P_n} \bar{\Pi}_{p+P_n\{k\}} \Bigg) \nonumber\\
\begin{picture}(100, 40)(0, 17)
\Line(20, 20)(50, 20)
\DashLine(50, 20)(80, 0){5}
\DashLine(50, 20)(80, 40){5}
\DashCArc(50,20)(30,-20,20){2}
\Text(20, 26)[]{$p$}
\Text(84, 48)[]{$k^{(1)}$}
\Text(84, -8)[]{$k^{(2n)}$}
\end{picture}
&\leftrightarrow& \frac{(-1)^n}{\hbar v^{2n-1}} \left( -2^{2n-1}m_w^2 - \left(1-2^{2n-1}\right)m_\eta^2 \right) \nonumber\\[20pt]
&\vdots& \nonumber\\
\begin{picture}(100, 40)(0, 17)
\Line(20,0)(50,20)
\Line(20,40)(50,20)
\DashLine(50, 20)(80, 0){5}
\DashLine(50, 20)(80, 40){5}
\DashCArc(50,20)(30,-20,20){2}
\Vertex(41.7,25.5){3}
\Text(20, 32)[]{$p$}
\Text(20, 10)[]{$q$}
\Text(84, 48)[]{$k^{(1)}$}
\Text(84, -8)[]{$k^{(2n)}$}
\end{picture}
&\leftrightarrow& \frac{(-1)^n}{\hbar v^{2n}} \left( -\bar{\Pi}_p \right) \nonumber\\[20pt]
\begin{picture}(100, 40)(0, 17)
\Line(20,0)(50,20)
\Line(20,40)(50,20)
\DashLine(50, 20)(80, 0){5}
\DashLine(50, 20)(80, 40){5}
\DashCArc(50,20)(30,-20,20){2}
\Vertex(41.7,14.5){3}
\Text(20, 32)[]{$p$}
\Text(20, 10)[]{$q$}
\Text(84, 48)[]{$k^{(1)}$}
\Text(84, -8)[]{$k^{(2n)}$}
\end{picture}
&\leftrightarrow& \frac{(-1)^n}{\hbar v^{2n}} \left( -\bar{\Pi}_q \right) \nonumber\\[20pt]
\begin{picture}(100, 40)(0, 17)
\Line(20,0)(50,20)
\Line(20,40)(50,20)
\DashLine(50, 20)(80, 0){5}
\DashLine(50, 20)(80, 40){5}
\DashCArc(50,20)(30,-20,20){2}
\Vertex(50,20){3}
\Text(20, 32)[]{$p$}
\Text(20, 10)[]{$q$}
\Text(84, 48)[]{$k^{(1)}$}
\Text(84, -8)[]{$k^{(2n)}$}
\end{picture}
&\leftrightarrow& \frac{(-1)^n}{\hbar v^{2n}}\Bigg( \sum_{P_1} \Pi_{p+P_1\{k\}} + \sum_{P_1} \Pi_{q+P_1\{k\}} + \nonumber\\
& & \phantom{\frac{(-1)^n}{\hbar v^{2n}}\Bigg(} -\sum_{P_2} \bar{\Pi}_{p+P_2\{k\}} - \sum_{P_2} \bar{\Pi}_{q+P_2\{k\}} + \ldots + \nonumber\\
& & \phantom{\frac{(-1)^n}{\hbar v^{2n}}\Bigg(} (-1)^n\sum_{P_{n-1}} \Pi_{p+P_{n-1}\{k\}} + (-1)^n\sum_{P_{n-1}} \Pi_{q+P_{n-1}\{k\}} + \nonumber\\
& & \phantom{\frac{(-1)^n}{\hbar v^{2n}}\Bigg(} \h(-1)^{n+1}\sum_{P_n} \bar{\Pi}_{p+P_n\{k\}} + \h(-1)^{n+1}\sum_{P_n} \bar{\Pi}_{q+P_n\{k\}} \Bigg) \nonumber\\
\begin{picture}(100, 40)(0, 17)
\Line(20,0)(50,20)
\Line(20,40)(50,20)
\DashLine(50, 20)(80, 0){5}
\DashLine(50, 20)(80, 40){5}
\DashCArc(50,20)(30,-20,20){2}
\Text(20, 32)[]{$p$}
\Text(20, 10)[]{$q$}
\Text(84, 48)[]{$k^{(1)}$}
\Text(84, -8)[]{$k^{(2n)}$}
\end{picture}
&\leftrightarrow& \frac{(-1)^n}{\hbar v^{2n}} \left( -2^{2n-1}m_w^2 + 2^{2n-1}m_\eta^2 \right) \nonumber\\[20pt]
&\vdots& \nonumber\\
\begin{picture}(100, 40)(0, 17)
\DashLine(50,20)(80,40){5}
\DashLine(50,20)(80,0){5}
\DashCArc(50,20)(30,50,310){2}
\Vertex(58.3,25.5){3}
\Text(84, 48)[]{$k^{(1)}$}
\Text(84, -8)[]{$k^{(2n)}$}
\end{picture}
&\leftrightarrow& \frac{(-1)^n}{\hbar v^{2n-2}} \Pi_{k^{(1)}} \nonumber\\[20pt]
\begin{picture}(100, 40)(0, 17)
\DashLine(50,20)(80,40){5}
\DashLine(50,20)(80,0){5}
\DashCArc(50,20)(30,50,310){2}
\Vertex(58.3,14.5){3}
\Text(84, 48)[]{$k^{(1)}$}
\Text(84, -8)[]{$k^{(2n)}$}
\end{picture}
&\leftrightarrow& \frac{(-1)^n}{\hbar v^{2n-2}} \Pi_{k^{(2n)}} \nonumber\\[20pt]
\begin{picture}(100, 40)(0, 17)
\DashLine(50,20)(80,40){5}
\DashLine(50,20)(80,0){5}
\DashCArc(50,20)(30,50,310){2}
\Vertex(50,20){3}
\Text(84, 48)[]{$k^{(1)}$}
\Text(84, -8)[]{$k^{(2n)}$}
\end{picture}
&\leftrightarrow& \frac{(-1)^n}{\hbar v^{2n-2}}\Bigg( -\sum_{P_2} \bar{\Pi}_{P_2\{k\}} + \ldots + \nonumber\\
& & \phantom{\frac{(-1)^n}{\hbar v^{2n-2}}\Bigg(} (-1)^n\sum_{P_{n-1}} \Pi_{P_{n-1}\{k\}} + \h(-1)^{n+1}\sum_{P_n} \bar{\Pi}_{P_n\{k\}} \Bigg) \nonumber\\
\begin{picture}(100, 40)(0, 17)
\DashLine(50,20)(80,40){5}
\DashLine(50,20)(80,0){5}
\DashCArc(50,20)(30,50,310){2}
\Text(84, 48)[]{$k^{(1)}$}
\Text(84, -8)[]{$k^{(2n)}$}
\end{picture}
&\leftrightarrow& \frac{(-1)^n}{\hbar v^{2n-2}} \left( -2^{2n-2}m_w^2 - \left(1-2^{2n-2}\right)m_\eta^2 \right) \nonumber\\[20pt]
&\vdots&
\eqa}

Having written the vertices in this form it is clear where the problem terms in a certain diagram come from. They come from a vertex with a dot in the center or two vertices connected by a line with two dots. That a dotted vertex is a source can immediately be seen from the vertex expressions above. A line with two dots and momentum $k$ flowing through it gets two $\Pi_k$'s in the numerator, from the vertices, and only one $\Pi_k$ in the denominator, from the propagator.

It is now easy to derive the following recursion relation, valid for $n\geq2$:
\vspace{-12pt}
{\allowdisplaybreaks\bqa
& & \begin{picture}(100, 40)(0, 17)
\DashLine(10,40)(40,20){5}
\DashLine(10,0)(40,20){5}
\DashLine(60,20)(90,40){5}
\DashLine(60,20)(90,0){5}
\Line(40,20)(60,20)
\Vertex(44,20){2}
\Vertex(56,20){2}
\DashCArc(60,20)(30,-28,28){2}
\Text(92,20)[l]{$2n-2$}
\end{picture} \hspace{40pt} +
\begin{picture}(100, 40)(0, 17)
\DashLine(10,40)(40,20){5}
\DashLine(10,0)(40,20){5}
\DashLine(10,26.7)(40,20){5}
\DashLine(10,13.3)(40,20){5}
\DashLine(60,20)(90,40){5}
\DashLine(60,20)(90,0){5}
\Line(40,20)(60,20)
\Vertex(44,20){2}
\Vertex(56,20){2}
\DashCArc(60,20)(30,-28,28){2}
\Text(92,20)[l]{$2n-4$}
\end{picture} \hspace{40pt} + \ldots + \nonumber\\[10pt]
& & \begin{picture}(100, 40)(0, 17)
\DashLine(10,40)(40,20){5}
\DashLine(10,0)(40,20){5}
\DashLine(60,20)(90,40){5}
\DashLine(60,20)(90,0){5}
\Line(40,20)(60,20)
\Vertex(44,20){2}
\Vertex(56,20){2}
\DashCArc(60,20)(30,-28,28){2}
\DashCArc(40,20)(30,152,208){2}
\Text(8,20)[r]{$e(n)$}
\Text(92,20)[l]{$2n-e(n)$}
\end{picture} \hspace{50pt} + \nonumber\\[10pt]
& & \begin{picture}(100, 40)(0, 17)
\DashLine(10,40)(40,20){5}
\DashLine(10,20)(40,20){5}
\DashLine(10,0)(40,20){5}
\DashLine(60,20)(90,40){5}
\DashLine(60,20)(90,0){5}
\DashLine(40,20)(60,20){5}
\Vertex(44,20){2}
\Vertex(56,20){2}
\DashCArc(60,20)(30,-28,28){2}
\Text(92,20)[l]{$2n-3$}
\end{picture} \hspace{40pt} +
\begin{picture}(100, 40)(0, 17)
\DashLine(10,40)(40,20){5}
\DashLine(10,30)(40,20){5}
\DashLine(10,20)(40,20){5}
\DashLine(10,10)(40,20){5}
\DashLine(10,0)(40,20){5}
\DashLine(60,20)(90,40){5}
\DashLine(60,20)(90,0){5}
\DashLine(40,20)(60,20){5}
\Vertex(44,20){2}
\Vertex(56,20){2}
\DashCArc(60,20)(30,-28,28){2}
\Text(92,20)[l]{$2n-5$}
\end{picture} \hspace{40pt} + \ldots + \nonumber\\[10pt]
& & \begin{picture}(100, 40)(0, 17)
\DashLine(10,40)(40,20){5}
\DashLine(10,0)(40,20){5}
\DashLine(60,20)(90,40){5}
\DashLine(60,20)(90,0){5}
\DashLine(40,20)(60,20){5}
\Vertex(44,20){2}
\Vertex(56,20){2}
\DashCArc(60,20)(30,-28,28){2}
\DashCArc(40,20)(30,152,208){2}
\Text(92,20)[l]{$e(n)+1$}
\Text(8,20)[r]{$2n-e(n)-1$}
\end{picture} \hspace{50pt} + \nonumber\\[20pt]
& & \begin{picture}(100, 40)(0, 17)
\DashLine(50,20)(80,40){5}
\DashLine(50,20)(80,0){5}
\Vertex(50,20){3}
\DashCArc(50,20)(30,50,310){2}
\Text(18,20)[r]{$2n$}
\end{picture} = 0
\eqa}\\[10pt]
In these diagrams it is understood that the outgoing legs should be connected in \emph{all} possible ways. $e(n)$ Is defined as:
\bq
e(n) \equiv \left\{ \begin{array}{ll}
n & \textrm{if $n$ is even} \\
n-1 & \textrm{if $n$ is odd}
\end{array} \right.
\eq
Notice that the third and fourth line in the recursion relation above are not there when $n=2$, these diagrams simply do not exist.

We also have the following two recursion relations:
{\allowdisplaybreaks\bqa
& & \begin{picture}(100, 40)(0, 17)
\DashLine(10,40)(40,20){5}
\DashLine(10,0)(40,20){5}
\DashLine(60,20)(90,40){5}
\DashLine(60,20)(90,0){5}
\Line(40,20)(60,20)
\Line(40,20)(90,50)
\Vertex(44,20){2}
\Vertex(56,20){2}
\DashCArc(60,20)(30,-28,28){2}
\Text(92,20)[l]{$2n-2$}
\end{picture} \hspace{40pt} + \ldots + \hspace{30pt}
\begin{picture}(100, 40)(0, 17)
\DashLine(10,40)(40,20){5}
\DashLine(10,0)(40,20){5}
\DashLine(60,20)(90,40){5}
\DashLine(60,20)(90,0){5}
\Line(40,20)(60,20)
\Line(40,20)(90,50)
\Vertex(44,20){2}
\Vertex(56,20){2}
\DashCArc(60,20)(30,-28,28){2}
\DashCArc(40,20)(30,152,208){2}
\Text(8,20)[r]{$e(n)$}
\Text(92,20)[l]{$2n-e(n)$}
\end{picture} \hspace{50pt} + \nonumber\\[20pt]
& & \begin{picture}(100, 40)(0, 17)
\DashLine(10,40)(40,20){5}
\DashLine(10,0)(40,20){5}
\DashLine(60,20)(90,40){5}
\DashLine(60,20)(90,0){5}
\Line(40,20)(60,20)
\Line(60,20)(90,50)
\Vertex(44,20){2}
\Vertex(56,20){2}
\DashCArc(60,20)(30,-28,28){2}
\Text(92,20)[l]{$2n-2$}
\end{picture} \hspace{40pt} + \ldots + \hspace{30pt}
\begin{picture}(100, 40)(0, 17)
\DashLine(10,40)(40,20){5}
\DashLine(10,0)(40,20){5}
\DashLine(60,20)(90,40){5}
\DashLine(60,20)(90,0){5}
\Line(40,20)(60,20)
\Line(60,20)(90,50)
\Vertex(44,20){2}
\Vertex(56,20){2}
\DashCArc(60,20)(30,-28,28){2}
\DashCArc(40,20)(30,152,208){2}
\Text(8,20)[r]{$e(n)$}
\Text(92,20)[l]{$2n-e(n)$}
\end{picture} \hspace{50pt} + \nonumber\\[20pt]
& & \begin{picture}(100, 40)(0, 17)
\DashLine(10,20)(40,20){5}
\DashLine(60,20)(90,40){5}
\DashLine(60,20)(90,0){5}
\DashLine(40,20)(60,20){5}
\Line(40,20)(90,50)
\Vertex(44,20){2}
\Vertex(56,20){2}
\DashCArc(60,20)(30,-28,28){2}
\Text(92,20)[l]{$2n-1$}
\end{picture} \hspace{40pt} + \ldots + \hspace{80pt}
\begin{picture}(100, 40)(0, 17)
\DashLine(10,40)(40,20){5}
\DashLine(10,0)(40,20){5}
\DashLine(60,20)(90,40){5}
\DashLine(60,20)(90,0){5}
\DashLine(40,20)(60,20){5}
\Line(40,20)(90,50)
\Vertex(44,20){2}
\Vertex(56,20){2}
\DashCArc(60,20)(30,-28,28){2}
\DashCArc(40,20)(30,152,208){2}
\Text(92,20)[l]{$e(n)+1$}
\Text(8,20)[r]{$2n-e(n)-1$}
\end{picture} \hspace{50pt} + \nonumber\\[20pt]
& & \begin{picture}(100, 40)(0, 17)
\DashLine(10,40)(40,20){5}
\DashLine(10,20)(40,20){5}
\DashLine(10,0)(40,20){5}
\DashLine(60,20)(90,40){5}
\DashLine(60,20)(90,0){5}
\DashLine(40,20)(60,20){5}
\Line(60,20)(90,50)
\Vertex(44,20){2}
\Vertex(56,20){2}
\DashCArc(60,20)(30,-28,28){2}
\Text(92,20)[l]{$2n-3$}
\end{picture} \hspace{40pt} + \ldots + \hspace{80pt}
\begin{picture}(100, 40)(0, 17)
\DashLine(10,40)(40,20){5}
\DashLine(10,0)(40,20){5}
\DashLine(60,20)(90,40){5}
\DashLine(60,20)(90,0){5}
\DashLine(40,20)(60,20){5}
\Line(60,20)(90,50)
\Vertex(44,20){2}
\Vertex(56,20){2}
\DashCArc(60,20)(30,-28,28){2}
\DashCArc(40,20)(30,152,208){2}
\Text(92,20)[l]{$e(n)+1$}
\Text(8,20)[r]{$2n-e(n)-1$}
\end{picture} \hspace{50pt} + \nonumber\\[30pt]
& & \begin{picture}(100, 40)(0, 17)
\DashLine(50,20)(80,40){5}
\DashLine(50,20)(80,0){5}
\Line(50,20)(80,50)
\DashLine(50,20)(20,40){5}
\Vertex(50,20){3}
\DashCArc(50,20)(30,156,316){2}
\Text(18,20)[r]{$2n$}
\end{picture} = 0
\eqa}\\[20pt]
In this recursion relation $n\geq2$.

{\allowdisplaybreaks\bqa
& & \begin{picture}(100, 40)(0, 17)
\DashLine(10,40)(40,20){5}
\DashLine(10,0)(40,20){5}
\DashLine(60,20)(90,40){5}
\DashLine(60,20)(90,0){5}
\Line(40,20)(60,20)
\Line(40,20)(90,60)
\Line(60,20)(90,50)
\Vertex(44,20){2}
\Vertex(56,20){2}
\DashCArc(60,20)(30,-28,28){2}
\Text(92,20)[l]{$2n-2$}
\end{picture} \hspace{40pt} + \ldots + \hspace{30pt}
\begin{picture}(100, 40)(0, 17)
\DashLine(10,40)(40,20){5}
\DashLine(10,0)(40,20){5}
\DashLine(60,20)(90,40){5}
\DashLine(60,20)(90,0){5}
\Line(40,20)(60,20)
\Line(40,20)(90,60)
\Line(60,20)(90,50)
\Vertex(44,20){2}
\Vertex(56,20){2}
\DashCArc(60,20)(30,-28,28){2}
\DashCArc(40,20)(30,152,208){2}
\Text(8,20)[r]{$e(n)$}
\Text(92,20)[l]{$2n-e(n)$}
\end{picture} \hspace{50pt} + \nonumber\\[30pt]
& & \begin{picture}(100, 40)(0, 17)
\DashLine(10,40)(40,20){5}
\DashLine(10,0)(40,20){5}
\DashLine(60,20)(90,40){5}
\DashLine(60,20)(90,0){5}
\Line(40,20)(60,20)
\Line(40,20)(90,50)
\Line(60,20)(90,60)
\Vertex(44,20){2}
\Vertex(56,20){2}
\DashCArc(60,20)(30,-28,28){2}
\Text(92,20)[l]{$2n-2$}
\end{picture} \hspace{40pt} + \ldots + \hspace{30pt}
\begin{picture}(100, 40)(0, 17)
\DashLine(10,40)(40,20){5}
\DashLine(10,0)(40,20){5}
\DashLine(60,20)(90,40){5}
\DashLine(60,20)(90,0){5}
\Line(40,20)(60,20)
\Line(40,20)(90,50)
\Line(60,20)(90,60)
\Vertex(44,20){2}
\Vertex(56,20){2}
\DashCArc(60,20)(30,-28,28){2}
\DashCArc(40,20)(30,152,208){2}
\Text(8,20)[r]{$e(n)$}
\Text(92,20)[l]{$2n-e(n)$}
\end{picture} \hspace{50pt} + \nonumber\\[30pt]
& & \begin{picture}(100, 40)(0, 17)
\DashLine(10,20)(40,20){5}
\DashLine(60,20)(90,40){5}
\DashLine(60,20)(90,0){5}
\DashLine(40,20)(60,20){5}
\Line(40,20)(90,60)
\Line(60,20)(90,50)
\Vertex(44,20){2}
\Vertex(56,20){2}
\DashCArc(60,20)(30,-28,28){2}
\Text(92,20)[l]{$2n-1$}
\end{picture} \hspace{40pt} + \ldots + \hspace{80pt}
\begin{picture}(100, 40)(0, 17)
\DashLine(10,40)(40,20){5}
\DashLine(10,0)(40,20){5}
\DashLine(60,20)(90,40){5}
\DashLine(60,20)(90,0){5}
\DashLine(40,20)(60,20){5}
\Line(40,20)(90,60)
\Line(60,20)(90,50)
\Vertex(44,20){2}
\Vertex(56,20){2}
\DashCArc(60,20)(30,-28,28){2}
\DashCArc(40,20)(30,152,208){2}
\Text(92,20)[l]{$e(n)+1$}
\Text(8,20)[r]{$2n-e(n)-1$}
\end{picture} \hspace{50pt} + \nonumber\\[30pt]
& & \begin{picture}(100, 40)(0, 17)
\DashLine(10,20)(40,20){5}
\DashLine(60,20)(90,40){5}
\DashLine(60,20)(90,0){5}
\DashLine(40,20)(60,20){5}
\Line(40,20)(90,50)
\Line(60,20)(90,60)
\Vertex(44,20){2}
\Vertex(56,20){2}
\DashCArc(60,20)(30,-28,28){2}
\Text(92,20)[l]{$2n-1$}
\end{picture} \hspace{40pt} + \ldots + \hspace{80pt}
\begin{picture}(100, 40)(0, 17)
\DashLine(10,40)(40,20){5}
\DashLine(10,0)(40,20){5}
\DashLine(60,20)(90,40){5}
\DashLine(60,20)(90,0){5}
\DashLine(40,20)(60,20){5}
\Line(40,20)(90,50)
\Line(60,20)(90,60)
\Vertex(44,20){2}
\Vertex(56,20){2}
\DashCArc(60,20)(30,-28,28){2}
\DashCArc(40,20)(30,152,208){2}
\Text(92,20)[l]{$e(n)+1$}
\Text(8,20)[r]{$2n-e(n)-1$}
\end{picture} \hspace{50pt} + \nonumber\\[40pt]
& & \begin{picture}(100, 40)(0, 17)
\DashLine(50,20)(80,40){5}
\DashLine(50,20)(80,0){5}
\Line(50,20)(80,50)
\Line(50,20)(80,60)
\DashLine(50,20)(20,40){5}
\Vertex(50,20){3}
\DashCArc(50,20)(30,156,316){2}
\Text(18,20)[r]{$2n$}
\end{picture} = 0
\eqa}\\[10pt]
In this recursion relation $n\geq1$. The first two lines are not there when $n=1$.

With these recursion relations it is easy to see that the problem terms always cancel in the complete set of diagrams for a certain Green's function. If somewhere in a diagram an internal line with two dots occurs, then at this same point in the diagram also the dotted vertex can occur. These diagrams then sum up to zero.

So finally we have proven that the problem terms cancel in the complete set of diagrams for a certain Green's function. If they cancel out anyway it is also correct to treat these problem terms like one would in the continuum. Of course one makes a mistake for each problem term, but these mistakes cancel out again in the complete set of diagrams. So there is nothing wrong with taking the continuum limit right from the start and doing a $d$-dimensional continuum calculation.

Notice that it is no problem to add the vertices coming from the potential $V$ and the Jacobian to this argument. These vertices give no problem terms themselves. They can be combined with the dotted vertices, but the problem terms always come from a \emph{clear}, \emph{separated} part of the diagram, either two vertices connected by a line with two dots, or a dotted vertex.

Now we know it is correct to take the continuum limit $\Delta\rightarrow0$ straightaway in the discrete Feynman rules (\ref{discrFeynmanrules}). If we do this it is easy to see that only the $\eta ww$- and $\eta\eta ww$-vertex do not vanish. All the other vertices go to zero, as can be seen from their expressions in (\ref{discrFeynmanrules}), but also, and much quicker, from (\ref{discrFeynman}), because they all have one or more factors of $\Delta$ in front when the exponentials are expanded.

So, finally we are left with the Feynman rules:
{\allowdisplaybreaks\bqa
\begin{picture}(100, 20)(0, 17)
\Line(20, 20)(80, 20)
\end{picture}
&\leftrightarrow& \frac{\hbar}{k^2+m_\eta^2} \nonumber\\
\begin{picture}(100, 20)(0, 17)
\DashLine(20, 20)(80, 20){5}
\end{picture}
&\leftrightarrow& \frac{\hbar}{k^2+m_w^2} \nonumber\\
\begin{picture}(100, 40)(0, 17)
\Line(20, 20)(50, 20)
\DashLine(50, 20)(80, 0){5}
\DashLine(50, 20)(80, 40){5}
\Text(80,30)[c]{$k_1$}
\Text(80,10)[c]{$k_2$}
\end{picture}
&\leftrightarrow& \frac{2}{\hbar v} k_1\cdot k_2 \nonumber\\
\begin{picture}(100, 40)(0, 17)
\Line(20,0)(50,20)
\Line(20,40)(50,20)
\DashLine(50,20)(80,40){5}
\DashLine(50,20)(80,0){5}
\Text(80,30)[c]{$k_1$}
\Text(80,10)[c]{$k_2$}
\end{picture}
&\leftrightarrow& \frac{2}{\hbar v^2} k_1\cdot k_2 \label{continuousFeynmanrules}
\eqa}\\
plus the Feynman rules coming from the potential $V$ and the Jacobian. These are exactly the Feynman rules that one would have read off from the continuum action:
\bqa
S &=& \int d^dx \left( \h\left(\nabla r\right)^2 + \h\left(\nabla w\right)^2 + {1\over v}\eta\left(\nabla w\right)^2 + {1\over2v^2}\eta^2\left(\nabla w\right)^2 + V\left(r\cos{w\over v},r\sin{w\over v}\right) \right) \nonumber\\
&=& \int d^dx \left( \h\left(\nabla r\right)^2 + \h{r^2\over v^2}\left(\nabla w\right)^2 + V\left(r\cos{w\over v},r\sin{w\over v}\right) \right) \label{continuumaction}
\eqa

Recapitulating the proof, we have done the following. The basis of our proof is the \emph{discrete} path integral in terms of the Cartesian fields $\f_1$ and $\f_2$. In this path integral on the lattice it is completely legitimate to transform to polar fields. After this transformation we get a very big, complicated action. Looking at the discrete vertex expressions we find how we can simplify these expressions, and how we can let them cancel against propagators in a certain diagram. We notice that these rules are exactly the same as in a $d$-dimensional continuum calculation. All the rules that we would use in a continuum calculation appear to be valid in a calculation on the lattice as well, except for one: rule (\ref{badrule}). The terms where we would need to use this rule can then be shown to cancel in the complete set of diagrams, by using the three recursion relations. So by using the incorrect rule (\ref{badrule}) we actually make a mistake, but all these mistakes cancel in the complete set of diagrams. Thus we know that \emph{all} the rules that we use in $d$-dimensional continuum calculation are \emph{also} valid in a correct, discrete calculation. This means we might as well take the continuum limit directly in the discrete Feynman rules (\ref{discrFeynman}). Then these Feynman rules simplify to (\ref{continuousFeynmanrules}), and we have proven that a $d$-dimensional continuum calculation with the action (\ref{continuumaction}) is \emph{correct}.

\subsection{An Example}

To see explicitly how the mechanism described in the previous section works we consider an example. Consider the 1-loop $\eta$-propagator. There are two types of diagrams (We do not include vertices from the potential $V$ and the Jacobian, because such vertices will never give problem terms.):
\vspace{-18pt}
\begin{center}
\begin{picture}(70,40)(0,18)
\Line(0,20)(20,20)
\DashCArc(35,20)(15,-180,180){3}
\Line(50,20)(70,20)
\end{picture} \quad and \quad
\begin{picture}(70,40)(0,18)
\Line(0,5)(70,5)
\DashCArc(35,20)(15,-90,270){3}
\end{picture}
\end{center}\vspace{10pt}
Here, dots should still be put on the lines or in the vertices. There are a lot of diagrams, but it is easy to see that there are only two diagrams that contain problem terms. These are:
\vspace{-17pt}
\bqa
\begin{picture}(70,40)(0,18)
\Line(0,20)(20,20)
\DashCArc(35,20)(15,-180,180){3}
\Line(50,20)(70,20)
\Vertex(48,27.5){2}
\Vertex(22,27.5){2}
\end{picture} &=& {1\over(2\pi)^d} \int_{-\pi/\Delta}^{\pi/\Delta} d^dk \; \left(-{1\over\hbar v}\right) \Pi_k \left(-{1\over\hbar v}\right) \Pi_k \frac{\hbar}{\Pi_k} \frac{\hbar}{\Pi_{p-k}} \nonumber\\
&=& {1\over v^2} {1\over(2\pi)^d} \int_{-\pi/\Delta}^{\pi/\Delta} d^dk \; \frac{\Pi_{p-k}}{\Pi_k} \nonumber\\
\begin{picture}(70,40)(0,18)
\Line(0,5)(70,5)
\DashCArc(35,20)(15,-90,270){3}
\Vertex(35,5){3}
\end{picture} &=& \h {1\over(2\pi)^d} \int_{-\pi/\Delta}^{\pi/\Delta} d^dk \; \left(-{1\over\hbar v^2}\right) \nonumber\\
& & \phantom{\h {1\over(2\pi)^d} \int_{-\pi/\Delta}^{\pi/\Delta} d^dk \;} \left( \h\left(\Pi_{p-k}+\Pi_{p+k}\right) + \h\left(\Pi_{p+k}+\Pi_{p-k}\right) \right) \frac{\hbar}{\Pi_k} \nonumber\\
&=& -{1\over v^2} {1\over(2\pi)^d} \int_{-\pi/\Delta}^{\pi/\Delta} d^dk \; \frac{\Pi_{p-k}}{\Pi_k}
\eqa
Indeed these diagrams cancel, as is guaranteed by the recursion relations derived in the previous section. The other diagrams, contributing to this $\eta$-propagator at 1-loop order, have their dots in other places, or have vertices without dots, that can also come from the potential $V$. These diagrams can never have a problem term. And thus the whole 1-loop propagator is free of problem terms, and the continuum limit could have been taken right from the start.

\subsection{The Jacobian and $w$-Loops}

In the previous sections it has become clear that it is allowed to work with the continuum Feynman rules (\ref{continuousFeynmanrules}), as the conjecture states. Together with these Feynman rules we have of course the rules from the arbitrary potential $V$, and the rules from the Jacobian. From the discrete calculation it is easy to see that the Jacobian can indeed be rewritten as:
\bqa
\prod_x r(x) &=& \prod_x \exp\left(-{1\over\hbar}\left(-\hbar\ln r(x)\right)\right) \nonumber\\
&=& \exp\left(-{1\over\hbar} \sum_x \left(-\hbar\ln r(x)\right)\right) \nonumber\\
&\rightarrow& \exp\left(-{1\over\hbar}{1\over\Delta^d} \int d^dx \left(-\hbar\ln r(x)\right)\right) \nonumber\\
&\rightarrow& \exp\left(-{1\over\hbar} \left[{1\over(2\pi)^d}\int d^dk\right] \int d^dx \left(-\hbar\ln r(x)\right)\right) \nonumber\\
&=& \exp\left(-{1\over\hbar} \left[{1\over(2\pi)^d}\int d^dk\right] \int d^dx \left(-\hbar \sum_{n=1}^\infty \frac{(-1)^{n+1}}{n} \left({\eta(x)\over v}\right)^n \right)\right) \;, \nonumber\\
\eqa
as the conjecture states. The Feynman rules coming from this Jacobian are:
\vspace{-17pt}
\bqa
\begin{picture}(100, 40)(0, 17)
\Line(20, 20)(50, 20)
\Vertex(50, 20){3}
\end{picture}
&\leftrightarrow& {1\over v} \; I \nonumber\\
\begin{picture}(100, 40)(0, 17)
\Line(20, 20)(80, 20)
\Vertex(50, 20){3}
\end{picture}
&\leftrightarrow& -{1\over v^2} \; I \nonumber\\
&\vdots& \nonumber\\
\begin{picture}(100, 40)(0, 17)
\Line(20, 20)(50, 20)
\Line(50, 20)(80, 40)
\Line(50, 20)(80, 0)
\DashCArc(50,20)(30,-30,30){2}
\Vertex(50, 20){3}
\Text(82,20)[l]{n}
\end{picture}
&\leftrightarrow& {(-1)^{n+1}(n-1)!\over v^n} \; I
\eqa\\[5pt]

We see that all these vertices give strange integrals $I$. In the case of the shifted toy model and the arctangent toy model we already saw that these integrals $I$ always cancelled against identical terms coming from $w$-loops. We shall now prove this in general.

Consider a $w$-loop with only $\eta ww$- and $\eta\eta ww$-vertices (as given in (\ref{continuousFeynmanrules})) on it. So the diagrams we are calculating can only have external $\eta$-legs. Such a diagram would look like:
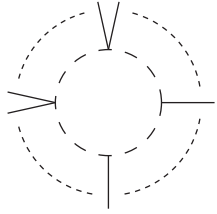
\begin{center}
\begin{picture}(80,80)(0,0)
\DashCArc(40,40)(20,0,360){5}
\Line(60,40)(80,40)
\Line(40,60)(36,78)
\Line(40,60)(44,78)
\Line(20,40)(2,44)
\Line(20,40)(2,36)
\Line(40,0)(40,20)
\DashCArc(40,40)(35,10,75){2}
\DashCArc(40,40)(35,105,165){2}
\DashCArc(40,40)(35,195,260){2}
\DashCArc(40,40)(35,280,350){2}
\end{picture}
\end{center}
Now these diagrams have a part which is going to cancel the $I$-integrals from the Jacobian. This part is exactly the worst divergent part of the diagrams above. To calculate this worst divergent part the masses and incoming momenta can be neglected.

In this case it is easy to write down the generating functional for such diagrams. This generating functional is defined as:
\vspace{-20pt}
\bq
Z(x) = \sum_{n=1}^\infty \quad
\SetScale{0.7}
\begin{picture}(56,56)(0,25)
\DashCArc(40,40)(20,0,360){5}
\Line(60,40)(80,40)
\Line(40,60)(36,78)
\Line(40,60)(44,78)
\Line(20,40)(2,44)
\Line(20,40)(2,36)
\Line(40,0)(40,20)
\DashCArc(40,40)(35,10,75){2}
\DashCArc(40,40)(35,105,165){2}
\DashCArc(40,40)(35,195,260){2}
\DashCArc(40,40)(35,280,350){2}
\end{picture} \quad \frac{x^n}{n!} \;,
\SetScale{1}
\eq\\[10pt]
where the diagram symbolizes \emph{all} 1-loop diagrams of this type with $n$ outgoing $\eta$-lines.

We denote the number of $\eta ww$-vertices in the $w$-loop by $n_3$ and the number of $\eta\eta ww$-vertices by $n_4$. Then the generating functional $Z(x)$ for diagrams of this type is given by:
\bqa
Z(x) &=& {1\over(2\pi)^d} \int d^dk \underset{n_3+n_4>0}{\sum_{n_3,n_4=0}^\infty} \left({1\over2!}\right)^{n_3} \left({1\over2!2!}\right)^{n_4} {1\over n_3!} {1\over n_4!} \left({2\over\hbar v}(-k^2)\right)^{n_3} \left({2\over\hbar v^2}(-k^2)\right)^{n_4} \nonumber\\
& & \phantom{{1\over(2\pi)^d} \int d^dk \underset{n_3+n_4>0}{\sum_{n_3,n_4=0}^\infty}} \left(\frac{\hbar}{k^2}\right)^{n_3+n_4} (2n_3+2n_4)!! (n_3+2n_4)! \frac{x^{n_3+2n_4}}{(n_3+2n_4)!} \nonumber\\
&=& -I \; \ln\left(1+{x\over v}\right) \nonumber\\
&=& I \; \sum_{n=1}^\infty \frac{(-1)^n(n-1)!}{v^n} \frac{x^n}{n!}
\eqa
So we can read off that a $w$-loop with $n$ outgoing $\eta$-lines has a worst divergent part given by:
\bq
\frac{(-1)^n(n-1)!}{v^n} \; I
\eq
This exactly cancels the vertices from the Jacobian. In any diagram, wherever a dotted vertex from the Jacobian with $n$ legs occurs, also a $w$-loop with $n$ outgoing legs can occur, and their part that contains the standard integral $I$ cancels!

\subsection{The Dimensional-Regularization Scheme}

In the case that one uses the dimensional-regularization scheme one has that:
\bq
{1\over(2\pi)^d} \int d^dk \; \frac{1}{(k^2)^m} = 0 \qquad \forall \; m \;,
\eq
which means that also our standard integral $I$ becomes zero:
\bq
I = 0 \;.
\eq
This means that in the dimensional-regularization scheme it becomes even easier to work with a path integral in polar field variables. In this case one can also completely forget about the Jacobian one gets from the transformation. Also one can ignore the integrals $I$ that are generated by $w$-loops.

In this paper we will keep everything general however, and not specify a regularization scheme.

\section{The One-Dimensional Case}

In section \ref{proofconj} we have proven the conjecture for a general model with two fields in $d$ space-time dimensions. This conjecture, which is promoted to a theorem by now, enables us to actually calculate things via the path integral in terms polar fields for any $d$-dimensional model. In a $d$-dimensional model the only analytical computations we can do in practice are continuum calculations. Analytical discrete calculations, i.e.\ calculations on the lattice, are in practice much too hard to do. That is why we have \emph{not} bothered to simplify the \emph{discrete} $d$-dimensional path integral, hoping to do a discrete calculation with this simplified form.

In \emph{one} dimension analytical discrete calculations \emph{are} sometimes possible (as we will see in the next section). Therefore it is convenient to have a reasonably simple, \emph{discrete} path integral in terms of polar fields for the case $d=1$. It is this path integral that we shall derive in this section.

By deriving this path integral we shall also make contact with the literature on quantum mechanical (i.e.\ 1-dimensional) path integrals in terms of polar fields, e.g.\ \cite{Lee,Edwards,Peak}.

Our starting point will again be the discrete Feynman rules (\ref{discrFeynman}), now specified to $d=1$ however. Also, for the sake of the argument, we will split up the vertices as given below. The one-dimensional Feynman rules are then:
{\allowdisplaybreaks\bqa
\begin{picture}(100, 20)(0, 17)
\Line(20, 20)(80, 20)
\end{picture}
&\leftrightarrow& \frac{\hbar}{{2\over\Delta^2}-{2\over\Delta^2}\cos{\Delta k}+m_\eta^2} \nonumber\\
\begin{picture}(100, 20)(0, 17)
\DashLine(20, 20)(80, 20){5}
\end{picture}
&\leftrightarrow& \frac{\hbar}{{2\over\Delta^2}-{2\over\Delta^2}\cos{\Delta k}+m_w^2} \nonumber\\
\begin{picture}(100, 40)(0, 17)
\Line(20, 20)(50, 20)
\DashLine(50, 20)(80, 0){5}
\DashLine(50, 20)(80, 40){5}
\DashCArc(50,20)(30,-20,20){2}
\Text(20, 26)[]{$p$}
\Text(84, 48)[]{$k^{(1)}$}
\Text(84, -8)[]{$k^{(2n)}$}
\end{picture}
&\leftrightarrow& \frac{2(-1)^n}{\hbar v^{2n-1}}{1\over\Delta^2} \left[ \left(e^{-i\Delta k^{(1)}}-1\right) \cdots \left(e^{-i\Delta k^{(2n)}}-1\right) \right] \nonumber\\[20pt]
\begin{picture}(100, 40)(0, 17)
\ArrowLine(20, 20)(50, 20)
\DashLine(50, 20)(80, 0){5}
\DashLine(50, 20)(80, 40){5}
\DashCArc(50,20)(30,-20,20){2}
\Text(20, 26)[]{$p$}
\Text(84, 48)[]{$k^{(1)}$}
\Text(84, -8)[]{$k^{(2n)}$}
\end{picture}
&\leftrightarrow& \frac{(-1)^n}{\hbar v^{2n-1}}{1\over\Delta^2} \left[ \left(e^{-i\Delta p}-1\right)\left(e^{-i\Delta k^{(1)}}-1\right) \cdots \left(e^{-i\Delta k^{(2n)}}\right) \right] \nonumber\\[20pt]
\begin{picture}(100, 40)(0, 17)
\Line(20,0)(50,20)
\Line(20,40)(50,20)
\DashLine(50, 20)(80, 0){5}
\DashLine(50, 20)(80, 40){5}
\DashCArc(50,20)(30,-20,20){2}
\Text(20, 32)[]{$p$}
\Text(20, 10)[]{$q$}
\Text(84, 48)[]{$k^{(1)}$}
\Text(84, -8)[]{$k^{(2n)}$}
\end{picture}
&\leftrightarrow& \frac{2(-1)^n}{\hbar v^{2n}}{1\over\Delta^2} \left[ \left(e^{-i\Delta k^{(1)}}-1\right) \cdots \left(e^{-i\Delta k^{(2n)}}-1\right) \right] \nonumber\\[20pt]
\begin{picture}(100, 40)(0, 17)
\Line(20,0)(50,20)
\ArrowLine(20,40)(50,20)
\DashLine(50, 20)(80, 0){5}
\DashLine(50, 20)(80, 40){5}
\DashCArc(50,20)(30,-20,20){2}
\Text(20, 32)[]{$p$}
\Text(20, 10)[]{$q$}
\Text(84, 48)[]{$k^{(1)}$}
\Text(84, -8)[]{$k^{(2n)}$}
\end{picture}
&\leftrightarrow& \frac{(-1)^n}{\hbar v^{2n}}{1\over\Delta^2} \left[ \left(e^{-i\Delta p}-1\right)\left(e^{-i\Delta k^{(1)}}-1\right) \cdots \left(e^{-i\Delta k^{(2n)}_1}-1\right) \right] \nonumber\\[20pt]
\begin{picture}(100, 40)(0, 17)
\DashLine(50,20)(80,40){5}
\DashLine(50,20)(80,0){5}
\DashCArc(50,20)(30,50,310){2}
\Text(84, 48)[]{$k^{(1)}$}
\Text(84, -8)[]{$k^{(2n)}$}
\end{picture}
&\leftrightarrow& \frac{(-1)^n}{\hbar v^{2n-2}}{1\over\Delta^2} \left[ \left(e^{-i\Delta k^{(1)}}-1\right) \cdots \left(e^{-i\Delta k^{(2n)}}-1\right) \right]
\eqa}\\

Looking at these vertex expressions we notice that only the vertices
\vspace{-17pt}
\bq \label{survivingvertices}
\begin{picture}(100, 40)(0,17)
\Line(20, 20)(50, 20)
\DashLine(50, 20)(80, 0){5}
\DashLine(50, 20)(80, 40){5}
\end{picture} \textrm{and}
\begin{picture}(100, 40)(0,17)
\Line(20,0)(50,20)
\Line(20,40)(50,20)
\DashLine(50, 20)(80, 0){5}
\DashLine(50, 20)(80, 40){5}
\end{picture}
\eq\\
have a finite continuum limit, \emph{all} the other vertices go to zero when $\Delta$ is sent to zero in the Feynman rules. First we are going to consider \emph{all} diagrams which have at least one of these vertices that vanish in the continuum limit. The only way these vertices can survive a continuum limit in a complete diagram is when there occur loops that give a $1/\Delta$.

First consider 1-loop diagrams. All 1-loop diagrams can be built from the vacuum diagram
\vspace{-17pt}
\bq
\begin{picture}(50, 40)(0,17)
\CArc(25,20)(15,0,360)
\end{picture}
\eq\\
By attaching legs we can build any 1-loop diagram from these. Having an $\eta$-line in the loop will never give a $1/\Delta$, no matter which vertices we use. If the whole loop is a $w$-line this loop can give $1/\Delta$'s. If we construct a diagram from this vacuum graph with at least one of the vertices that go to zero in the continuum limit, one can verify easily that either the whole diagram goes to zero in the continuum limit \emph{or} diagrams cancel among each other in the complete set of graphs for a certain process, such that the whole process is zero.

The same thing can now be done on 2-loop level. Here we can construct all diagrams from the vacuum graphs
\vspace{-17pt}
\bq
\begin{picture}(100, 40)(0,17)
\CArc(35,20)(15,0,360)
\CArc(65,20)(15,-180,180)
\end{picture},
\begin{picture}(100, 40)(0,17)
\Line(30,20)(50,20)
\Line(50,20)(70,20)
\CArc(50,20)(20,0,360)
\end{picture} \textrm{and} \quad
\begin{picture}(100, 40)(0,17)
\Line(40,20)(60,20)
\CArc(25,20)(15,0,360)
\CArc(75,20)(15,-180,180)
\end{picture}
\eq\\
One can see that the only diagrams surviving the continuum limit and containing at least one of the vertices that vanish in the continuum limit can be constructed from the following vacuum graphs by \emph{only} attaching lines with the vertices (\ref{survivingvertices}), because these vertices do not give additional powers of $\Delta$.
\vspace{-17pt}
\bq \label{twoloopvacgraphs}
\begin{picture}(100, 40)(0,17)
\DashCArc(35,20)(15,0,360){5}
\DashCArc(65,20)(15,-180,180){5}
\end{picture},
\begin{picture}(100, 40)(0,17)
\DashCArc(35,20)(15,0,360){5}
\CArc(65,20)(15,-180,90)
\ArrowArc(65,20)(15,90,180)
\end{picture},
\begin{picture}(100, 40)(0,17)
\Line(30,20)(50,20)
\ArrowLine(50,20)(70,20)
\DashCArc(50,20)(20,0,360){5}
\end{picture} \textrm{and}
\begin{picture}(100, 40)(0,17)
\ArrowLine(50,20)(30,20)
\ArrowLine(50,20)(70,20)
\DashCArc(50,20)(20,0,360){5}
\end{picture}
\eq\\
For the vacuum graphs we have the following expressions, excluding vertex constants and symmetry factors. Only the discrete loop integration is done and the worst behavior in $\Delta$ is kept.
\vspace{-17pt}
\bqa
\begin{picture}(100, 40)(0,17)
\DashCArc(35,20)(15,0,360){5}
\DashCArc(65,20)(15,-180,180){5}
\end{picture} &=& \frac{\hbar^2}{\Delta^2} \nonumber\\
\begin{picture}(100, 40)(0,17)
\DashCArc(35,20)(15,0,360){5}
\CArc(65,20)(15,-180,90)
\ArrowArc(65,20)(15,90,180)
\end{picture} &=& -\h\frac{\hbar^2}{\Delta} \nonumber\\
\begin{picture}(100, 40)(0,17)
\Line(30,20)(50,20)
\ArrowLine(50,20)(70,20)
\DashCArc(50,20)(20,0,360){5}
\end{picture} &=& -\h\frac{\hbar^3}{\Delta} \nonumber\\
\begin{picture}(100, 40)(0,17)
\ArrowLine(50,20)(30,20)
\ArrowLine(50,20)(70,20)
\DashCArc(50,20)(20,0,360){5}
\end{picture} &=& \frac{\hbar^3}{\Delta^2}
\eqa\\

We now construct all 1PI diagrams from these vacuum graphs by attaching lines with the vertices (\ref{survivingvertices}). Because we can only attach lines with these vertices we can only get external $\eta$-lines. We shall now calculate the generating functional of all the diagrams that can be constructed in this way:
\vspace{-17pt}
\bq
Z(x) = \sum_{n=0}^\infty 
\begin{picture}(60, 40)(0,17)
\Line(30,20)(55,40)
\Line(30,20)(57.7,36)
\Line(30,20)(55,0)
\DashCArc(30,20)(25,-34,26){2}
\Text(57,20)[l]{n}
\GCirc(30,20){15}{0.5}
\end{picture} \quad {1\over n!} x^n
\eq\\

Now this $Z$, for the first vacuum graph, is given by:
\bqa
Z &=& {1\over8}{\hbar^2\over\Delta^2} \sum_{m_3,m_4,n_3,n_4=0}^\infty \left( {\Delta^2\over\hbar v^2} + {2\Delta^2\over\hbar v^3}x + {2\Delta^2\over\hbar v^4}\h x^2 \right) \hbar^{m_3+m_4+n_3+n_4} \nonumber\\
& & \phantom{\h{\hbar^2\over\Delta^2}{1\over4} \sum_{m_3,m_4,n_3,n_4=0}^\infty} \left(-\frac{2}{\hbar v}\right)^{m_3+n_3} \left(-\frac{2}{\hbar v^2}\right)^{m_4+n_4} \left({1\over2!}\right)^{m_3+n_3} \left({1\over2!2!}\right)^{m_4+n_4} \nonumber\\
& & \phantom{\h{\hbar^2\over\Delta^2}{1\over4} \sum_{m_3,m_4,n_3,n_4=0}^\infty} {1\over m_3!} {1\over n_3!} {1\over m_4!} {1\over n_4!} (2m_3+2m_4)!! (2n_3+2n_4)!! x^{m_3+n_3+2n_4+2m_4} \nonumber\\
&=& {1\over8}{\hbar\over v^2} \frac{1}{\left(1+{x\over v}\right)^2} \nonumber\\
&=& {\hbar\over8} \sum_{n=0}^\infty \frac{(-1)^n (n+1)!}{v^{n+2}} {1\over n!} x^n \label{genfunctvacblob}
\eqa
Here $1/8$ is the symmetry factor of the vacuum graph, $m_3$ and $m_4$ denote respectively the number of $\eta ww$- and $\eta\eta ww$-vertices in the left loop and $n_3$ and $n_4$ denote the number of $\eta ww$- and $\eta\eta ww$-vertices in the right loop. Now we can read off
\vspace{-17pt}
\bq \label{Leeterm}
\begin{picture}(60, 40)(0,17)
\Line(30,20)(55,40)
\Line(30,20)(57.7,36)
\Line(30,20)(55,0)
\DashCArc(30,20)(25,-34,26){2}
\Text(57,20)[l]{n}
\GCirc(30,20){15}{0.5}
\end{picture}  \quad = {\hbar\over8} \frac{(-1)^n (n+1)!}{v^{n+2}}
\eq\\[5pt]
for the first vacuum graph.

Also the contributions from the three other vacuum blobs can be constructed in the same way. Their generating functions appear to cancel each other. The reason for this shall become clear below. For now the \emph{only} 2-loop contribution we get is (\ref{Leeterm}).

The $n$-leg diagrams that we find in (\ref{Leeterm}) are exactly the vertices one would get from a term
\bq \label{Leeterm2}
-{\hbar^2\over8}{1\over r_i^2}
\eq
in the action. This can easily be seen by substituting $r=v+\eta$ in this term and expanding it in $\eta$. This means, up to 2-loop level, one can discard the vertices that go to zero in the continuum limit and replace them by the vertices from (\ref{Leeterm2}). In the action this means one is left with
\bq
S = \Delta \sum_{i=0}^{N-1} \left( \h\frac{(r_{i+1}-r_i)^2}{\Delta^2} + \h {r_i^2\over v^2}\frac{(w_{i+1}-w_i)^2}{\Delta^2} - \frac{\hbar^2}{8r_i^2} \right) \;.
\eq

Disregarding three- and higher-loop level we have now proven that our one-dimensional discrete path integral $P$ is equal to:
\bqa
P &=& \int_{-\infty}^{\infty} dr_0 \ldots dr_{N-1} \; r_0 \ldots r_{N-1} \; \int_{-\infty}^{\infty} dw_0 \ldots dw_{N-1} \; O \nonumber\\
& & \qquad \exp\Bigg( -\frac{1}{\hbar} \Delta \sum_{i=0}^{N-1} \bigg( \h\frac{(r_{i+1}-r_i)^2}{\Delta^2} + \h {r_i^2\over v^2}\frac{(w_{i+1}-w_i)^2}{\Delta^2} - \frac{\hbar^2}{8r_i^2} + \nonumber\\
& & \phantom{\qquad \exp\Bigg( -\frac{1}{\hbar} \Delta \sum_{i=0}^{N-1} \bigg(} V\left(r_i\cos{w_i\over v},r_i\sin{w_i\over v}\right) \bigg) \Bigg) \;. \label{1dimpathintLee}
\eqa

This is a form that one can also find in the literature. This same path integral is derived by Lee \cite{Lee} in chapter 19, formula (19.49). Also Edwards et al.\ \cite{Edwards} and Peak et al.\ \cite{Peak} find a term (\ref{Leeterm2}). However they start with the discrete path integral in terms of Cartesian fields, transform to polar fields and actually perform the angular integration. Only then they find the term (\ref{Leeterm2}). We have presented a more general proof of this term here, like Lee \cite{Lee}.

Up to now we have \emph{not} proven that at three- and higher-loop level there are diagrams, containing at least on of the vertices that vanish in the continuum limit, that can \emph{not} be built also from vertices from the term (\ref{Leeterm2}). We shall not prove this in this paper. In this paper we are mostly interested in the transformation to polar fields in $d$-dimensional models, and the conjecture needed to compute via polar fields in these models has already been fully proven. It should be clear however that, to have agreement with the literature, the path integral (\ref{1dimpathintLee}) is correct, up to all orders. So, although we cannot prove it at this point, there are \emph{no} diagrams at three- and higher-loop order that cannot also be constructed with only the term (\ref{Leeterm2}).

\subsection{An Alternative Derivation}

Another way to derive the discrete one-dimensional path integral (\ref{1dimpathintLee}) is by first using the conjecture. So one computes all diagrams in a $d$-dimensional way in the continuum, with the simple Feynman rules from the continuum action, then one lets $d\rightarrow1$. To obtain the discrete version of these diagrams one has to know what the difference is between calculating in the continuum and calculating in the discrete. This difference, we know, comes from the problem terms. If we formulate the (continuum) $\eta ww$- and $\eta\eta ww$-vertex with the dots, as we did in section \ref{proofconj}, then we have a clear source of problem terms. Because in this case we \emph{only} have the $\eta ww$- and $\eta\eta ww$-vertex we also \emph{only} have the recursion relation:
\bqa
& & \begin{picture}(100, 40)(0, 17)
\DashLine(10,20)(90,20){5}
\Line(40,20)(10,50)
\Line(60,20)(90,50)
\Vertex(44,20){2}
\Vertex(56,20){2}
\end{picture} +
\begin{picture}(100, 40)(0, 17)
\DashLine(10,20)(90,20){5}
\Line(40,20)(90,50)
\Line(60,20)(10,50)
\Vertex(44,20){2}
\Vertex(56,20){2}
\end{picture} +
\begin{picture}(100, 40)(0, 17)
\DashLine(10,20)(90,20){5}
\Line(50,20)(90,50)
\Line(50,20)(10,50)
\Vertex(50,20){3}
\end{picture} = 0 \label{specialrecrel}
\eqa\\
This recursion relation ensures that all problem terms from a $w$-line with two dots cancel against the dotted part of the $\eta\eta ww$-vertex. So the problem terms from $w$-lines are never going to give a difference between a discrete and continuum calculation. All we have to do is find the problem terms coming from $\eta$-lines with two dots.

Now, as in the previous (partial) derivation of (\ref{1dimpathintLee}) we can build all diagrams from the vacuum graphs. At 1-loop order there is \emph{no} difference between a continuum and discrete calculation. At 2-loop order the only problem terms come from the vacuum graph
\vspace{-17pt}
\bq \label{dotgraph}
\begin{picture}(100, 40)(0,17)
\Line(30,20)(70,20)
\DashCArc(50,20)(20,0,360){5}
\Vertex(40,20){3}
\Vertex(60,20){3}
\end{picture}
\eq\\
Now we can understand why, in the previous derivation of (\ref{1dimpathintLee}), the generating functionals from the last three vacuum graphs in (\ref{twoloopvacgraphs}) cancelled. Only the first graph in (\ref{twoloopvacgraphs}) corresponds to the vacuum graph above. This correspondence can be seen by pinching the dotted $\eta$-line in the vacuum graph above. The last three vacuum graphs in (\ref{twoloopvacgraphs}) correspond to problem terms from dotted $w$-lines or a dotted $\eta\eta ww$-vertex. These cancel among each other because of the recursion relation (\ref{specialrecrel}).

Now the difference between a continuum and discrete calculation of the graph (\ref{dotgraph}) can be calculated. Also the generating functional of diagrams where we connect any number of $\eta$-legs via the $\eta ww$- and $\eta\eta ww$-vertices can be calculated. The result of this generating functional is given by (\ref{genfunctvacblob}). In this way we find that, to compensate for the differences that we get by doing a discrete instead of a continuum calculation, we have to introduce the term (\ref{Leeterm2}) in the action again.

Also in this way of deriving (\ref{1dimpathintLee}) we do not know how to show that three- and higher-loop diagrams give no new differences, but this is not important for the main result of this paper.

For a nice illustration of the strictly one-dimensional path integral in terms of polar fields (\ref{1dimpathintLee}) we refer to \cite{vanKessel}.

\section{Conclusion}

We have presented a way to rewrite a $d$-dimensional Euclidean path integral, in terms of two scalar fields $\f_1$ and $\f_2$, in terms of the corresponding polar fields $r$ and $w$. Our first step was to introduce a conjecture that stated how to perform this transformation. This conjecture states that (\ref{conjecture}) is the correct way to transform. After this we showed, by doing explicit calculations for two toy models, that the conjecture is indeed correct for these toy models up to some order in perturbation theory. Finally we gave a general proof of the conjecture, based on a completely discrete (i.e.\ lattice) calculation. To make contact with the literature on the transformation to polar fields in the case of quantum mechanics, we also specify our calculation to this case ($d=1$), and find agreement.

In a forthcoming paper we will use the conjecture in actual calculations on the Euclidean $N=2$ linear sigma model.

\appendix

\section{Standard Integrals}\label{appstandint}

Throughout this paper we use the following standard integrals:
\bqa
& & I\left(q_1,m_1,q_2,m_2,\ldots,q_n,m_n\right) \equiv \nonumber\\
& & \qquad {1\over(2\pi)^d} \int d^dk \; \frac{1}{\left(k+q_1\right)^2+m_1^2} \frac{1}{\left(k+q_2\right)^2+m_2^2} \cdots \frac{1}{\left(k+q_n\right)^2+m_n^2}
\eqa
\bqa
D_{m_1m_2m_3} &\equiv& {1\over(2\pi)^{2d}} \int d^dk \; d^dl \; \frac{1}{k^2+m_1^2} \; \frac{1}{l^2+m_2^2} \; \frac{1}{(k-l)^2+m_3^2} \\
B_{m_1m_2m_3} &\equiv& {1\over(2\pi)^{2d}} \int d^dk \; d^dl \; \frac{1}{(k^2+m_1^2)^2} \; \frac{1}{l^2+m_2^2} \; \frac{1}{(k-l)^2+m_3^2} \\
A_m(x) &\equiv& {1\over(2\pi)^d} \int d^dk \; e^{ik\cdot x} \; \frac{1}{k^2+m^2} \\
C_{m_1m_2}(x) &\equiv& {1\over(2\pi)^d} \int d^dk \; e^{ik\cdot x} \; \frac{1}{k^2+m_1^2} \; \frac{1}{k^2+m_2^2} \\
D_{m_1m_2m_3}(x) &\equiv& {1\over(2\pi)^{2d}} \int d^dk \; d^dl \; e^{ik\cdot x} \; \frac{1}{k^2+m_1^2} \; \frac{1}{l^2+m_2^2} \; \frac{1}{(k-l)^2+m_3^2} \\
B_{m_1m_2m_3}(x) &\equiv& {1\over(2\pi)^{2d}} \int d^dk \; d^dl \; e^{ik\cdot x} \; \frac{1}{(k^2+m_1^2)^2} \; \frac{1}{l^2+m_2^2} \; \frac{1}{(k-l)^2+m_3^2} \\
I &\equiv& {1\over(2\pi)^d} \int d^dk
\eqa

\end{document}